\documentclass[twocolumn,tighten]{aastex62}
\pdfoutput=1
\usepackage{amsmath,amstext,amssymb}
\usepackage[T1]{fontenc}
\usepackage{apjfonts} 
\usepackage[figure,figure*]{hypcap}
\usepackage{comment}
\usepackage{subfigure}
\usepackage[ampersand]{easylist}
\usepackage{multirow}
\usepackage{booktabs}
\usepackage{wrapfig}
\usepackage[document]{ragged2e}
\usepackage{epstopdf}

\makeatletter
\renewcommand\subsubsection{\@startsection{subsubsection}{3}{\z@}%
                                     {-3.25ex\@plus -1ex \@minus -.2ex}%
                                     {1.5ex \@plus .2ex}%
                                     {\normalfont\normalsize\bfseries}}

\makeatother

\shorttitle{X-ray Reverberation of \oneh} 
\shortauthors{Frederick et al.}

\newcommand{\oneh}{1H 1934-063~} 
\newcommand{\onehns}{1H 1934-063}
\newcommand{\xmm}{\textit{XMM-Newton}~}
\newcommand{\xmmns}{\textit{XMM-Newton}}
\newcommand{\nstr}{\textit{NuSTAR}~}
\newcommand{\nstrns}{\textit{NuSTAR}}
\newcommand{\code}[1]{\texttt{#1}} 
\newcommand{\pn}{\textit{EPIC-pn}~}

\accepted{\today}

\begin{document}
\justify
\title{X-ray Reverberation Mapping and Dramatic Variability of Seyfert 1 Galaxy 1H 1934-063} 

\author[0000-0001-9676-730X]{Sara Frederick}
\email{sfrederick@astro.umd.edu}
\affiliation{Department of Astronomy, University of Maryland, College Park, MD 20742, USA} 

\author[0000-0003-0172-0854]{Erin Kara}
\affiliation{Department of Astronomy, University of Maryland, College Park, MD 20742, USA} 
\affiliation{Joint Space-Science Institute, University of Maryland, College Park, MD 20742, USA}

\author[0000-0002-1510-4860]{Christopher Reynolds}
\author[0000-0003-2532-7379]{Ciro Pinto}
\author[0000-0002-9378-4072]{Andrew Fabian}
\affiliation{Institute of Astronomy, University of Cambridge, Madingley Road, Cambridge CB3 0HA, UK}

\begin{abstract} 
A fraction of active galactic nuclei (AGN) exhibit dramatic variability, which is observed on timescales down to minutes in the X-ray band. 
We introduce the case study of 1H 1934-063 (z = 0.0102), a Narrow-line Seyfert 1 (NLS1)  among the brightest and most variable AGN ever observed with \textit{XMM-Newton}. This work includes spectral and temporal analyses of a concurrent \textit{XMM-Newton} and \textit{NuSTAR} 2015 observation lasting 130 kiloseconds, during which the X-ray source exhibited a steep (factor of $\sim$6) plummet and subsequent full recovery of flux level, accompanied by deviation from a single log-normal flux distribution. 
We rule out Compton-thin obscuration as the cause for this dramatic variability observed even at \textit{NuSTAR} energies. In order to constrain coronal geometry, dynamics, and emission/absorption processes, we compare detailed spectral fitting with Fourier-based timing analysis. Similar to other well-studied, highly variable Seyfert 1s, this AGN is X-ray bright and displays strong reflection features. We find a narrower broad iron line component compared to most Seyfert 1s, and  constrain black hole spin to be < 0.1, one of the lowest yet discovered for such systems. Combined spectral and timing results are consistent with a dramatic change in the continuum on timescales as short as a few kiloseconds dictating the nature of this variability. 
We also discover a Fe-K time lag, measuring a delay of 20 seconds between relativistically-blurred reflection off the inner accretion flow and the hard X-ray continuum emission.

\end{abstract}

\keywords{accretion, accretion disks --- black hole physics --- galaxies: active --- galaxies: individual (1H 1934-063) --- galaxies: Seyfert --- relativistic processes}

\section{Introduction}
\label{intro}
An AGN can directly influence the evolution of its host galaxy via the sustained release of energy powered by accretion onto a central supermassive black hole (SMBH). 
An optically-thick, geometrically-thin ($h/r \sim 0.1$) accretion disk forms around the SMBH down to the innermost stable circular orbit (ISCO), due to the gravitational infall of accreted matter balanced by the outward transport of angular momentum. 
The majority of the sustained high-energy emission released from AGN host galaxies originates from the immediate vicinity of the black hole. 
 AGN flux is observed to vary stochastically in all wavebands and on timescales down to less than a day, implying light-hour-sized X-ray emitting regions \citep{davison1975increase,1975ApJ...199L.139W,1976ApJ...207L.159I}. 
 The complex processes underlying variability in each waveband remain hotly debated in the literature; origins such as variable line-of-sight obscuration or more fundamental changes such as accretion rate evolution or abrupt magnetic reconfiguration would all lead to variations in AGN luminosity. 

 The most variable AGN, taken together, are a compelling wellspring of interesting accretion-related phenomena. 
Focusing on unique case studies that probe extremes of the AGN mass ($10^6  \leq M_\text{BH} \leq10^{10} M_\odot$) and luminosity ($10^{42}\leq L_\text{bol} \leq 10^{46} \text{ erg s}^{-1}$) distributions, or exhibit paradigm-challenging variability properties, can help bridge gaps between MHD accretion disk theory and  observations. 

A typical AGN X-ray spectrum is dominated by a power-law continuum\textemdash the signature of inverse-Compton scattering of disk photons by an optically-thin X-ray corona comprised of hot electrons \citep{sunyaev1980comptonization,haardt1993x}. 
Magnetic fields threading the ionized accretion disc are thought to both confine and accelerate these energetic particles to power the compact, dynamic corona \citep{galeev1979structured,haardt1991two,merloni2001accretion,reis2013size,wilkins2015driving}. 
On average, AGN are measured with X-ray power-law photon indices $\Gamma\approx1.9-2.0$, values consistent with thermal Comptonization in a sphere with optical depth of order unity and temperature of 100 keV \citep{nandra1994ginga,brandt1997comparison}. 
Sometimes a high-energy exponential cutoff of this power-law continuum is observed ($F(E) \propto E ^{-\Gamma} e^{-E/E_{cut}}$), from which this fiducial temperature of the corona is estimated \citep{jourdain1992sigma,gondek1996average,fabian2015properties}. 

Coronal X-rays irradiating the accretion disk induce fluorescence, the most prominent emission being from iron \citep{george1991x}. A number of features arise from the resulting reflection spectrum, representing distinct structures at various distances along the accretion disk, and elucidating the interplay between absorption, line emission, and scattering at these high energies \citep{fabian2010x}. 
When reflection occurs at the inner edge of the accretion disk, this fluorescent emission can appear relativistically smeared, as observed in the broadened Fe-K$\alpha$ line. 
The ``Compton hump'', first observed with the Ginga X-ray satellite \citep{piro1990x,matsuoka1990x,pounds1990x}, is evident above 10 keV and interpreted as Compton reflection of continuum photons by the accretion disk (and as far out as the surrounding molecular torus). 

In more than 50\% of Seyfert 1s, a soft-X-ray excess in flux below $\sim$1 keV is observed in addition to the power law continuum at varying strengths, depending somewhat on the level of absorption affecting the observation \citep{singh1985observation,arnaud1985exosat,turner1989exosat}. 
This soft excess can be modeled physically as a reflection component made up of lines from the inner ionized disk which have been relativistically broadened beyond identification \citep{gierlinski2004soft,
crummy2006explanation}, an inner Compton upscattering region of the disk powered by the accretion flow \citep{done2012intrinsic}, an additional steep power law component due to the launching of a jet \citep{kataoka2007probing,chatterjee2009disk}, or a number of other diverse ideas that have been proposed; however, its origin remains largely mysterious and best-fit models are largely degenerate \citep{page2004xmm,porquet2004xmm,piconcelli2005xmm,schurch2006failed,dewangan2007investigation,lohfink2012black}. 
The importance of understanding this feature is evident from its dominating influence over the crucial spectral features in the highest signal-to-noise X-ray band.

Evidence for the soft excess temporally lagging behind the hard continuum was first uncovered in 1H0707-495 \citep{2009Natur.459..540F}. Measurements of energy-dependent lags led to the discovery of broad iron line lags, first in NGC 4151 \citep{2012MNRAS.422..129Z} and subsequently in several other sources \citep{2013ApJ...764L...9C,2013MNRAS.434.1129K,2014MNRAS.439L..26K,2014MNRAS.445...56K,2015MNRAS.446..737K,kara2016global}. All of these (high-frequency) lags are interpreted as reverberation of the X-ray continuum off the accretion disk. 
The rapid variability of the power law continuum in the hard band supports that this emission is close to the central black hole.
Low-frequency hard lags in AGN may be due to fluctuations propagating inwardly through the accretion disk, as is likely the case in black hole X-ray binaries \citep{kotov2001x}. 

\oneh (z = 0.0102, \citealp{rodriguez2008swift}; $M_{\rm BH} = 3\times10^6~M_{\odot}$, \citealp{rodriguez2000visible}), also known as IGR J19378-0617, SS 442, or IRAS 19348-0619, is a radio-quiet Narrow-line Seyfert 1 (NLS1; \citealp{nagao2001narrow}) which was ranked as 7th in 10-20 ks excess variance among 161 AGN comprising a subset of the Catalog of AGN in the \xmm Archive (CAIXAvar; \citealp{ponti2012caixa}). NLS1s are characterized by the presence of narrower broad Balmer lines ($\Delta v < 2000$ km/s), weak [OIII], and strong optical Fe II lines in what otherwise appears as a Seyfert 1 spectrum \citep{osterbrock1985spectra,goodrich1989spectropolarimetry}, indicating they tend to host super-Eddington-accreting, lower-mass SMBHs  \citep{boroson1992emission,pounds1995re,komossa2006radio,jin2015strong}. 
The BH mass of \oneh is estimated from FWHM H$\beta$ to be $M_{\rm BH} = 3\times10^6~M_{\odot}$ \citep{rodriguez2000visible,malizia2008first}. 
 Despite its similarities to archetypal X-ray AGN also ranking highly in variability by CAIXA, little has been published on this source. 

\oneh was reported by two INTEGRAL/IBIS catalogs \citep{molkov2004hard,bird2007third}, the MAXI 37-month source catalog \citep{hiroi201337}, as well as the Swift BAT 70-month catalog \citep{baumgartner201370}. 
The AGN is hosted by a low-Galactic-latitude face-on Sb-type Seyfert 1 galaxy. 
 It is a radio-quiet single unresolved radio source, with integrated 20-cm flux density $S_{\text{1.4 \rm{GHz}}}$ = 42.2 mJy = 4.2$\times 10^{-25}$ erg s$^{-1}$ cm$^{-2}$ Hz$^{-1}$ in the NRAO VLA Sky Survey \citep{condon1998nrao}.
A 12 ks observation of \oneh is described by \citealp{panessa2011narrow} to display the most prominent soft excess in a sample of 14 NLS1 spectra. They also report a broad-band photon index of 2.58 $\pm$ 0.02.
\oneh has a Galactic absorbing column of $N_{\rm H} = 1.48\times10^{21}$ cm$^{-2}$ measured by \citealp{dickey1990hi}, and $N_{\rm H} = 1.06\times10^{21}$ cm$^{-2}$ measured by \citealp{kalberla2005leiden}.

The fact that this AGN is X-ray bright and rapidly variable has allowed it to be studied at a wide range of timescales and to a great degree of detail. 
We present the first spectral and temporal analyses of a concurrent $\sim$130 ks \xmm and \nstr observation of \oneh (PI: E. Kara). 
This paper is organized as follows: Section~\ref{obs} summarizes the observations and data reduction and extraction; Section~\ref{spec} details the time-resolved spectral analysis and Section~\ref{tim} details the temporal analysis. Section~\ref{discuss} presents our physical interpretation of these results. 

\section{Observations}
\label{obs}
\oneh was observed concurrently with \xmm and \nstr on 2015 October 1-3 with exposures of 140 and 130 ks, respectively. 
Details of the observations, including the flux in the hard and soft bands, are presented in Table~\ref{tab_obs}. We reduced the data, as well as all subsequent time interval and energy-resolved extractions, following the standard processing procedures, outlined in the following sections.

\begin{table*}[!ht]
\centering
\begin{tabular}{l l l l l l}
\toprule
\multicolumn{6}{c}{Overview of \xmm and \nstr Observations} \\
\midrule
Satellite & OBSID & Start Date/Time & Exposure (s) & 0.5$-$2 keV Flux  (ergs cm$^{-2}$ s$^{-1}$)& 2$-$10 keV Flux (ergs cm$^{-2}$ s$^{-1}$) \\
\hline 
\hline
\xmm & 0761870201 &2015-10-01 15:31:18  & 141400 & $ 2.2\times 10^{-11}$  & $ 2.1 \times 10^{-11}$ \\  
 \nstr & 60101003002 & 2015-10-01 17:46:08 & 132000 & \nodata & \nodata  \\ 
\bottomrule
\end{tabular}
\caption{Properties of the \xmm and \nstr observations analyzed in this work, including observation time, net exposure, and flux measured with the best-fit model from this work. For more details, see Section~\ref{sec_softspc}.}
\label{tab_obs}
\end{table*}

\subsection{\xmm}
This work utilizes \textit{XMM-Newton}'s \pn camera, as well as the Optical Monitor (\textit{OM})  \citep{struder2001european}. 
The Reflection Grating Spectrometer \textit{RGS}, well-suited for high resolution spectroscopy below 3 keV, also collected data during this observation, detailed analysis of which we defer to a later paper. 
We do not consider data from the 2 \textit{EPIC} Metal Oxide Semi-conductor (\textit{MOS}) cameras, as they have lower count rate and are more susceptible to pile-up than the other detectors. 

The \xmm observation of \oneh analyzed here (Observation ID: 0761870201) began on 2015 October 1 with the \pn camera in small window and optical thin filter 1 mode, and the \textit{OM} in image mode with the \textit{UVW2} filter. 
The data were reduced using the \xmm Science Analysis System (SAS) v16.0 \citep{gabriel2004xmm}. 
\pn source products were extracted from a circular region of radius 30 arcseconds, and background products were extracted from a source-free 30-arcsecond-radius region.
We adopted CCD event patterns 0 to 4, corresponding to single- and double-pixel events.
To maximize signal-to-noise, we screened for periods of high background and therefore omit the first 15 ks of data. 
While determining a best-fit spectral model, we omit \xmm counts above 7 keV. At those energies, photon pile-up alters the shape of the spectrum for different annular extraction regions, which effectively mask the central pixels corresponding to the peak of the 2D PSF (where higher count rates increase the likelihood of more than one X-ray photon arriving in a single camera pixel before readout). We instead selected a circular extraction region (thereby maximizing the signal-to-noise), and during analysis masked the \xmm spectral data where we observed divergence of the \textit{EPIC-pn} and \nstr in their highest overlapping energies. We confirm that excising the innermost source emission rather than excluding the 7-10 keV data segment does not significantly alter the results reported in the spectral analysis. 
Spectra were binned using the \code{GRPPHA} command such that there are no less than 20 counts in each bin, to enable the use of $\chi^2$ statistics. 
Response files were created using the \code{RMFGEN} and \code{ARFGEN} commands, (corresponding to the redistribution matrices and ancillary response files, respectively). 
The \textit{OM} spectra were reduced using the \code{OMICHAIN} command, which performs aperture photometry of the sources present in \textit{OM} image data. In order to perform analysis on the \textit{OM} data, we retrieved the spectral response file for the \textit{UVW2} filter from the ESA webpage\footnote{\url{http://xmm2.esac.esa.int/external/xmm_sw_cal/calib/om_files.shtml}}. 
We used \xmm \pn calibration database files updated as of May 2017.

\subsection{\nstr}
\nstr is complementary to \textit{XMM-Newton}'s sensitivity range in the soft X-ray band by providing constraints on the reflection-dominated 10$-$80 keV band, where it is 100 times more sensitive than its predecessors \citep{harrison2013nuclear}. 
It consists of 2 semiconductor arrays, Focal Plane Modules A and B (FPMA/FPMB), at the focal point of the first focusing hard X-ray mirror placed on an astrophysical X-ray satellite. 

\nstr observed \oneh concurrently with \xmm for exposure times of roughly 70 ks per instrument (Observation ID: 60101003002). 
The data were processed with \nstr Data Analysis Software (NuSTARDAS) v1.6.0 using the \code{NUPIPELINE} task.
Source and background products were extracted from circular regions with radii of 120 arcseconds,
 (there is not a significant change in photon index for different extraction regions). 
The effective on-source time after SAA filtering was 68 ks per FPM (136 ks = 97\% of total integrated on-source time.) 
Spectra were binned to over-sample the instrumental resolution by a factor of 3 and to a minimum signal to noise ratio of 5 in each spectral channel, after background subtraction. 
 We utilized the most up-to-date available calibration files (February 2017).

\section{Analysis \& Results}
\label{sec_results}
\subsection{X-ray Photometry}
\label{lcs}
\begin{figure}[h!] 
\centering
\includegraphics[scale=0.45]{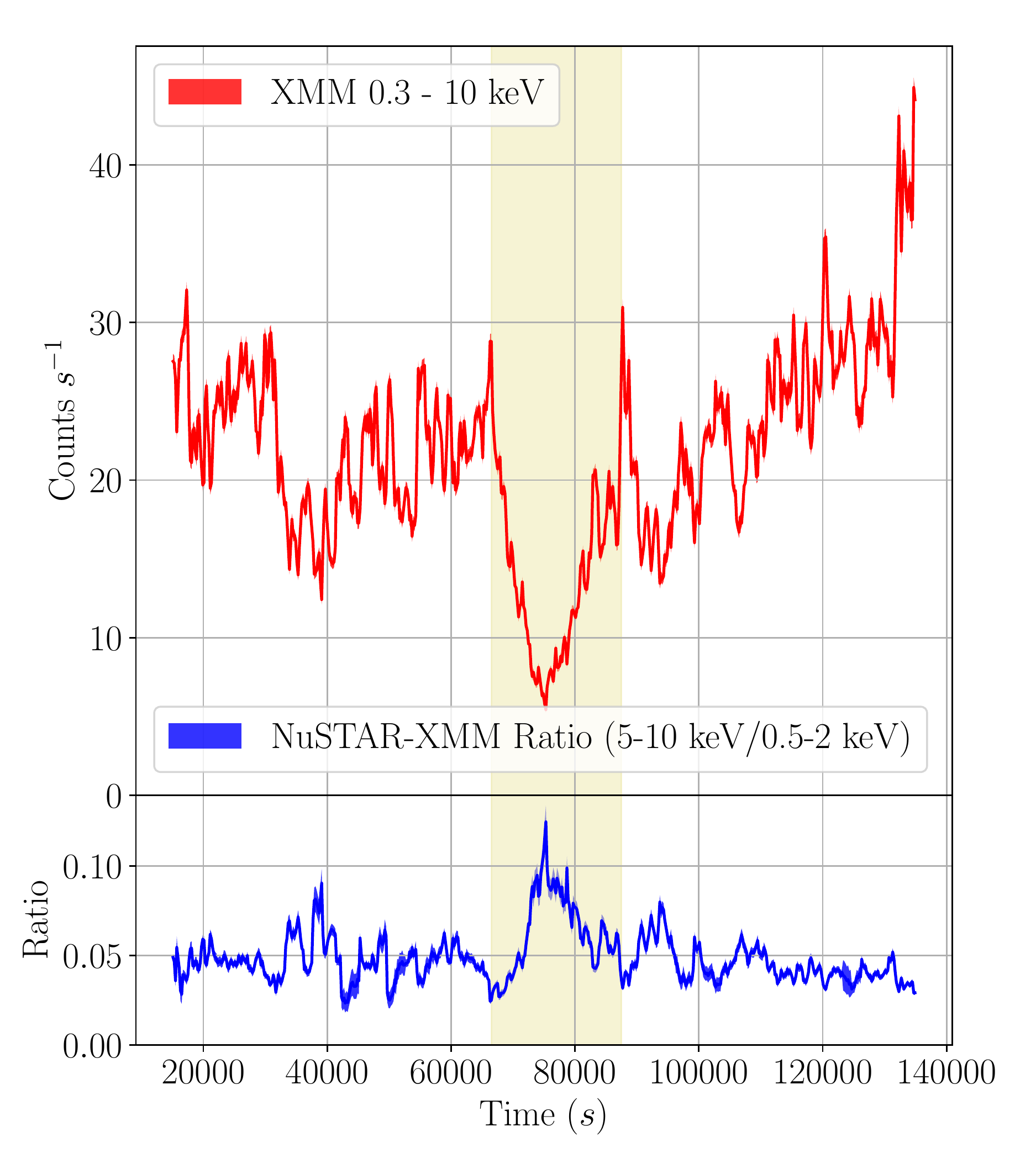}
\caption{Top panel: \xmm single orbit 0.3$-$10 keV light curve of \oneh (bin size 200 seconds). Note the high-amplitude variability and highlighted deep drop in flux during the central 20 ks of the observation. Error bars are shown but of comparable size to the line width. Lower panel: Hardness ratio light curve (\nstr 5$-$10 keV light curve divided by \xmm 0.5$-$2 keV light curve) indicate that the emission becomes harder during the central 20 ks of the observation.}
\label{bblc}
\end{figure}

A cursory glance at the photometric data from the \xmm observation of \oneh reveals that it is extremely time-variable and X-ray bright (evident in the broad-band light curve in Figure~\ref{bblc} reaching 40 counts s$^{-1}$). 

The hardness ratio light curve of \oneh is shown in the lower panel of Figure \ref{bblc}. It is evident that spectral variability is taking place during the entire observation, with the central 20 ks of the observation showing a major change where a strong hardening of the emission occurs. 
This agrees with the softer-when-brighter paradigm that has been well-established with months-long observations of variable Seyferts \citep{treves1990x,leighly1996x}. A deficit in soft photons would also agree qualitatively with extra absorption during the dip state (e.g. \citealp{sanfrutos2013size}).  
Secondary peaks in the hardness ratio roughly correspond to other shorter-duration dips in the light curve.

\begin{figure}[h!] 
\centering
\includegraphics[scale=0.4]{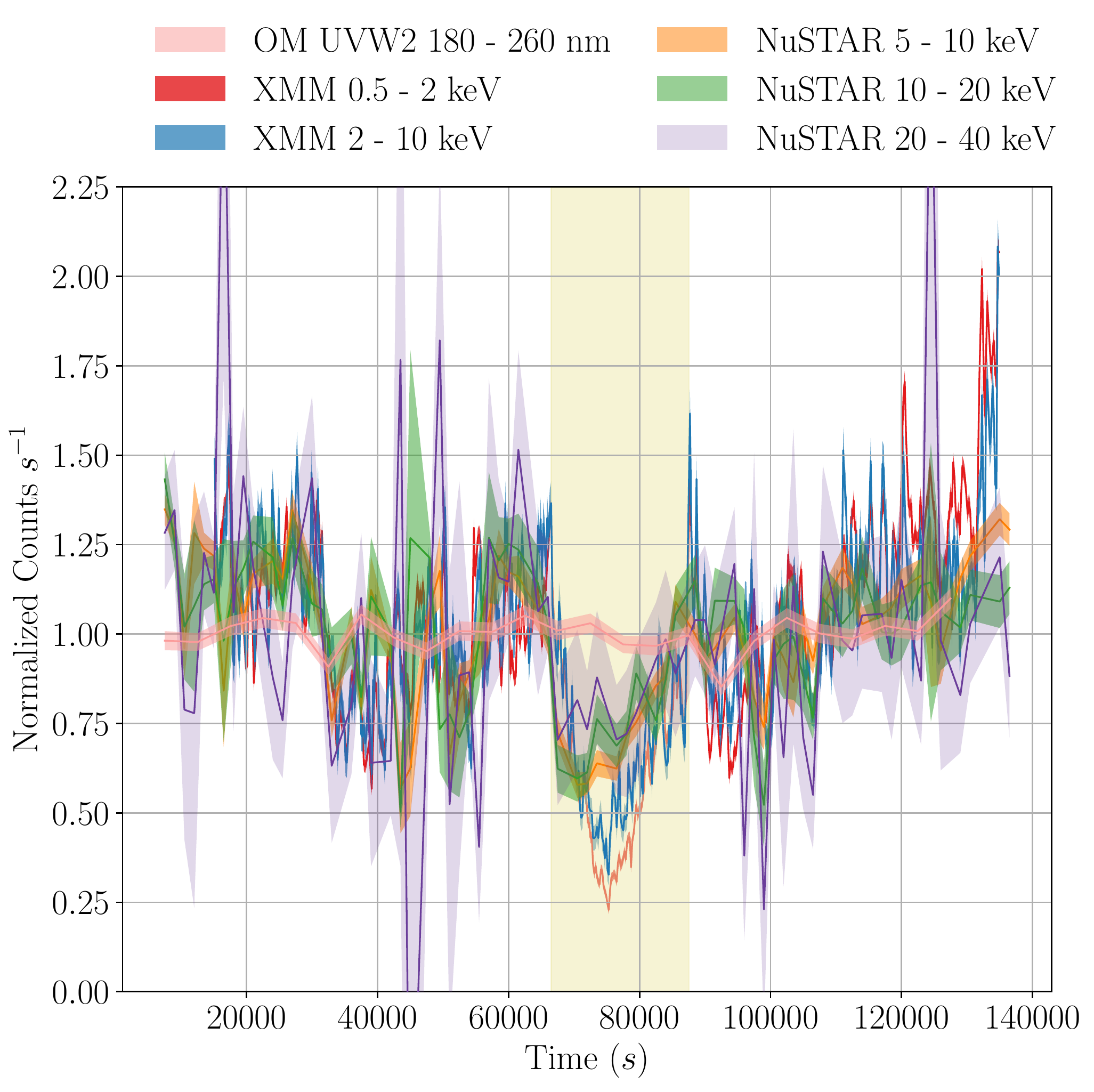}
\caption{\xmm \textit{EPIC-pn}, \textit{OM}, and \nstr energy-resolved light curves (bin size 200, 5000, and 1500 seconds respectively). Higher energy bins are represented by bluer colors, and softer by redder colors. The low flux event can be seen to occur within the central 20 ks of the observation for both X-ray telescopes up to 20 keV, but not in the optical. We combine data from the 2 \nstr instruments (FPMA and FPMB). Each light curve is normalized to its average count rate to display the fractional flux decrease in each energy band.}
\label{normlc}
\end{figure}

Both the \xmm and \nstr X-ray light curves, shown in Figure~\ref{normlc} for several energy bands, show \oneh undergoing dramatic variability in the form of a symmetric decrease and subsequent recovery of flux spanning the central 20 ks of both observations. There is no such clear change in flux state observed in the \textit{OM} \textit{UVW2}-filter data.
We have normalized these light curves to the average flux level to compare the fractional flux decrease for each energy band directly; we observe that higher energy emission seen with NuSTAR appears to be slightly less affected by this low flux event, as it drops by a factor of $\sim$2 compared to the \xmm decrease by a factor of $\sim$6. 
The fact that the \nstr light curve shows this dip may be an indicator that this is an intrinsic change in X-ray emission. In Section~\ref{spc_bme}, we present time-resolved spectroscopy testing if this intrinsic variability model can describe the data well. 
Although the low flux event in the \nstr band appears by-eye to slightly precede the soft \xmm band dip, the same temporal offset is not seen in the XMM hard light curve in the NuSTAR energy band (5-10 keV). This apparent hard lead is likely the effect of random fluctuations, as well as binning over orbital gaps (regular occultations due to \textit{NuSTAR}'s 600-km nearly-circular orbit, \citealp{harrison2013nuclear}). 

\begin{figure}[h!] 
\centering
\includegraphics[scale=0.35]{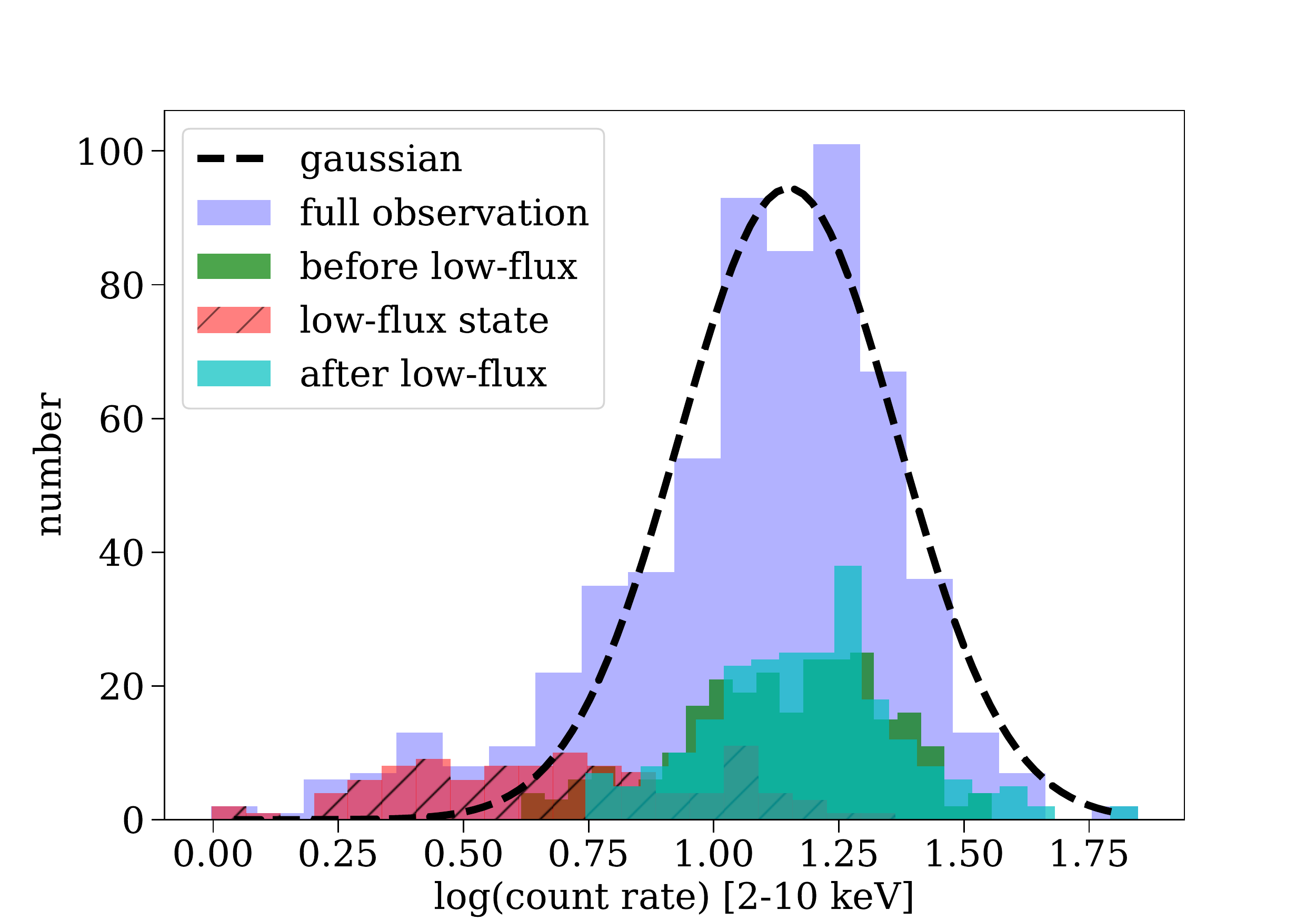}
\caption{We bin the light curve for each part of the observation (purple=full 120 ks, green=initial high flux state, pink/lined=during the 20-ks low-flux state, and blue=the full recovery of flux) and see that during the high-flux states, \oneh displays the lognormal flux distribution typical of most AGN, with the low-flux state skewing the average away from lognormal and comprising the tail of this distribution.} 
\label{hist}
\end{figure}

A lognormal distribution of fluxes is commonly observed in the long-term stochastic X-ray variability of AGN \citep{gaskell2004lognormal}. 
It is apparent from Figure~\ref{hist} that the count rate of the lognormal flux histogram for this observation deviates from the average during the 20 ks of the low-flux state, though it is less clear whether the overall shape of the distribution undergoes a change. 
This may indicate an extreme variability event taking place on very fast timescales; therefore, we investigate the data during each of these stages (pre-dip high flux state, low-flux state, and post-dip recovery of flux) in Sections \ref{spc_bme} and \ref{tim_bme}. 

\subsection{Spectroscopy}
\label{spec}
For the spectral analysis outlined in this section, we used \code{XSPEC} version 12.9.1a \citep{1996ASPC..101...17A} and assess best-fit models utilizing $\chi^2$ statistics. 
Uncertainties are quoted at 90\% confidence intervals unless otherwise stated. 

We use the cosmic abundances of \citealp{wilms2000absorption} and the photoelectric absorption
cross sections of \citealp{verner1996atomic}.
We also assume the following cosmology for luminosity calculations: $H_0$ = 70 km s$^{-1}$ Mpc$^{-1}$,  $\Omega_\Lambda$= 0.73 and $\Omega_M$ = 0.27. 
The photon index $ \Gamma$ is defined by flux density $F(E) \propto E ^{-\Gamma}$. 
When comparing spectra from different instruments, a multiplicative constant is introduced in order to account for discrepancies in the absolute flux calibrations. 
The \nstr instruments are not sufficiently calibrated below 3 keV and above 78 keV, therefore we do not fit to spectra above and below these thresholds. Similarly, we omit the \textit{EPIC-pn} data below 0.3 keV where \xmm is not reliably calibrated. 

\subsubsection{Modeling the Relativistic Iron Line}
\label{sec_fe}
In order to spectroscopically characterize the strong relativistic reflection in \onehns, we first investigate the \nstr spectra and the \pn spectra above 2 keV. This is to exclude the large soft-excess in this source, which dominates the signal-to-noise at soft X-ray energies (0.3$-$2 keV), 
and therefore avoid any skewing effects it may introduce to the reflection model. This has the added benefit of not allowing the least understood spectral feature to dominate the initial fit (see Section~\ref{intro} for details). 
Instead the description of the relativistic reflection in this Section relies primarily on the prominent Fe-K$\alpha$ line (6$-$7 keV).

\paragraph*{\textbf{Time-integrated Spectroscopy\\}}
\label{spc_int}

\begin{figure}[h!] 
\centering
\includegraphics[scale=0.24]{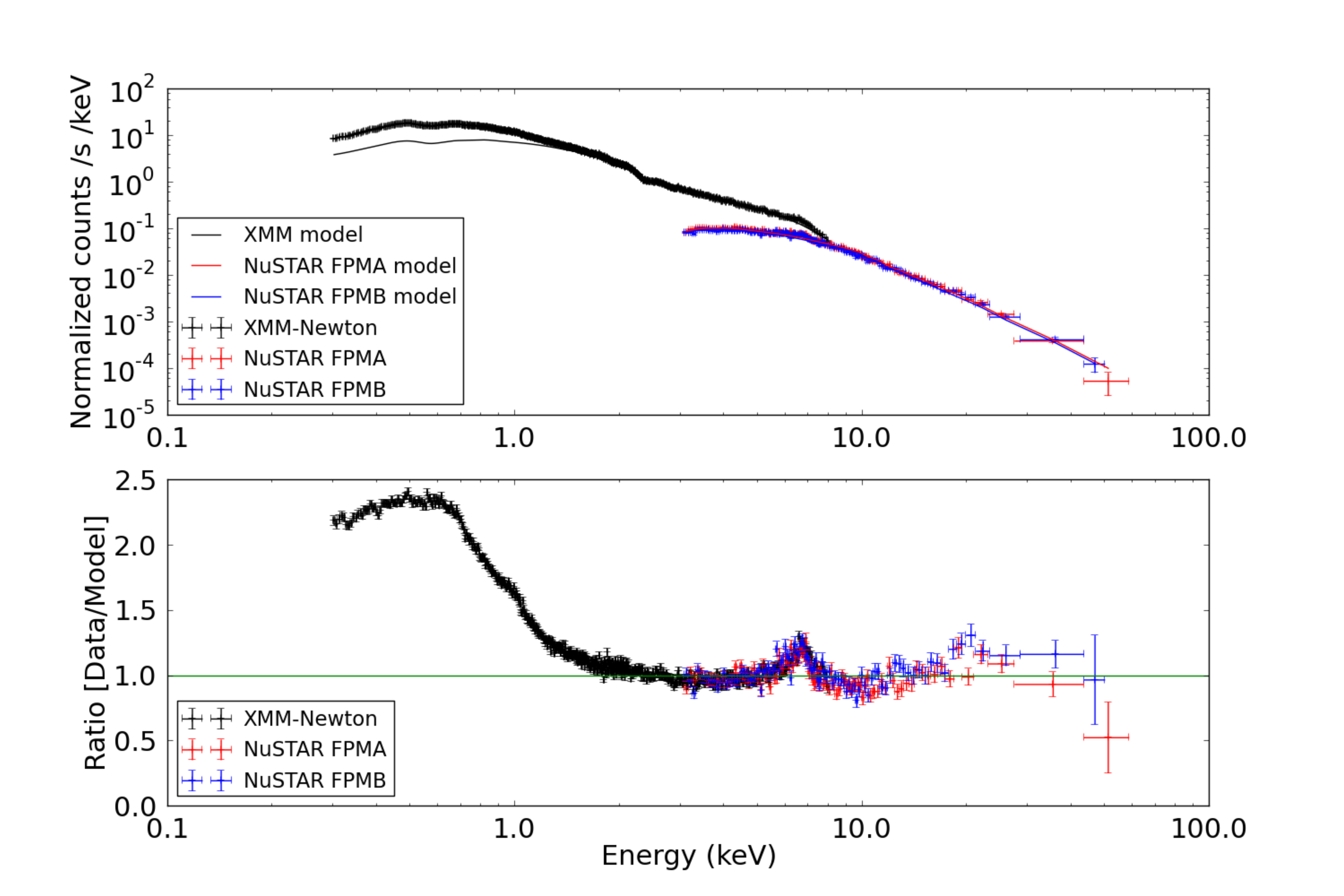}
\caption{\xmm (black) and \nstr (red and blue representing both \nstr focal plane modules) spectra fit to a simple absorbed power law, excluding energies below 2 keV so that the fit is not dominated by the soft excess. Spectra throughout this work have been binned slightly for visualization purposes. The power law ($\Gamma \sim 2.1$) is evident in the time-integrated spectrum (upper panel); the soft excess below 2 keV, iron line (6$-$7 keV), and Compton hump above 10 keV appear in the data-to-model ratio residuals  (lower panel).}
\label{spectra}
\end{figure}

We begin by modeling the \xmm (2$-$8 keV) and \nstr (3$-$60 keV) time-integrated spectra of \oneh using a power law with Galactic absorption, shown superimposed on the data in the upper panel of Figure~\ref{spectra}. 
Upon first inspection, the residuals (the lower panel of Figure~\ref{spectra}) display strong reflection features\textemdash a broad iron line, Compton hump, and dominant soft excess can be seen by eye. These features are explored in detail in Section~\ref{spc_bme} and Section~\ref{sec_softspc}. 

We observe the power law photon index to be slightly discrepant between \xmm and \nstrns, likely due to differences in instrument calibration as well as the telescopes sampling different energy ranges. 
We therefore allow the photon index $\Gamma$ to vary freely between the two data sets, although for the best-fit model the \xmm and \nstr power law photon indices are consistent within error.

The Galactic absorption is represented by the photoelectric absorption model \code{tbnew} v2.3, which accounts for absorption by the solid state of iron in the form of dust grains \citep{wilms2000absorption}. 
Because the measured Galactic absorption is discrepant between $N_{\rm{H}} = 1.5\times10^{21}$ cm$^{-2}$ \citep{dickey1990hi} and $N_{\rm{H}} = 1.06\times10^{21}$ cm$^{-2}$ \citep{kalberla2005leiden}, we allow this parameter to vary and measure $N_{\rm{H}} = 1.26\times10^{21}$ cm$^{-2}$. 
This discrepancy in measured Galactic column is likely due to the low Galactic latitude of this source; the line-of-sight to \oneh grazes the outskirts of the Galactic disk in IR observations with the Two Micron All-Sky Survey (2MASS; \citealp{cutri2003irsa}). 
We initially allow the abundances for Fe and O to vary; however, the fit is not sensitive to these parameters, so we therefore fix them both to Solar abundances. 

We begin to explore the iron line profile with the following models, all subject to Galactic absorption: 
\begin{easylist}[enumerate]
& a power law continuum (\code{po})
& ionized reflection in the Newtonian limit (\code{xillver}) 
& ionized and relativistically blurred reflection (\code{relxill}). 
\end{easylist}

\begin{figure}[ht!] 
\centering
\includegraphics[scale=0.58]{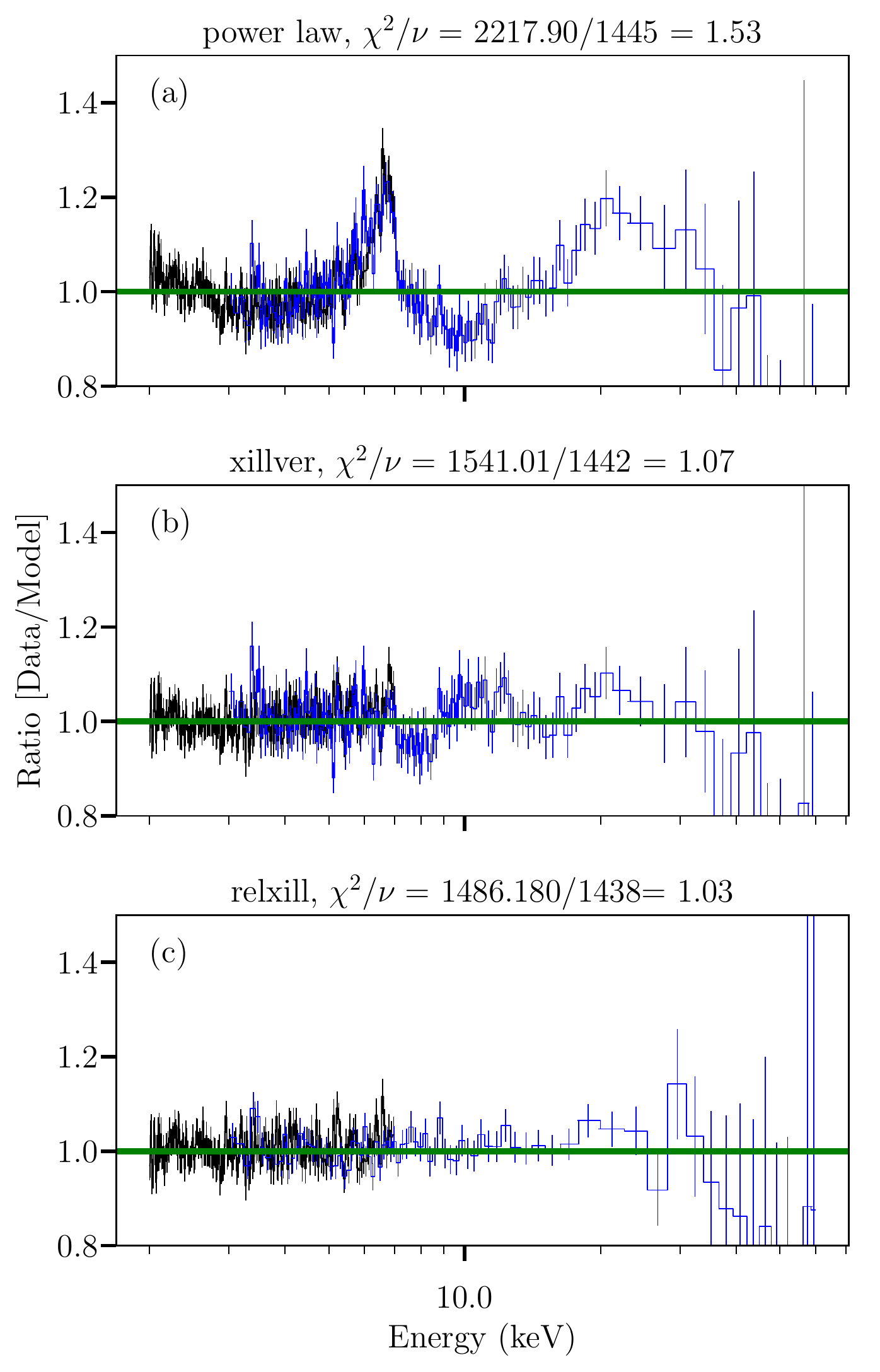}
\caption{Increasingly complex models describing the \xmm and \nstr data between 2 and 60 keV, and their respective goodness-of-fit values. The data are well described with an ionized, relativistically broadened reflection model. All are convolved with Galactic absorption treated with \code{tbnew}. The differences between the multiplicative models are as follows: a) \code{power law}, b)  \code{xillver}\textemdash ionized reflection, c) \code{relxill} relativistically blurred reflection. Data from both \nstr FPMs have been grouped for visual purposes} 
\label{line_tests}
\end{figure}

 The model fits can be seen to iteratively improve ($\chi^2$ listed in Figure~\ref{line_tests}) with each subsequent physical addition, with a clear preference for the relativistically smeared ionized reflection family of models (\code{relxill}). 
Other model flavors of \code{relxill} give equally good fits  \citep{dauser2014role,garcia2014improved,dauser2016normalizing}.

For the spectroscopic analysis to follow, we utilize the \code{relxill} model in the lamp-post geometry, with the X-ray emitting region approximated as a point source above a thin disk and along the axis of azimuthal symmetry (\code{relxilllp}). This model is not as physical as those accounting for the emissivity structure of a geometrically extended corona or finite disk thickness \citep{taylor2018exploring}. It also does not account for a multitude of phenomena which may alter the X-ray spectrum in important ways, such as re-emission of optically thick matter at the ISCO or coronal emission from the underside of the accretion disk \citep{niedzwiecki2018lamppost}, but has been shown to describe observed accretion disk irradiation profiles well \citep{dauser2010broad,garcia2011x,garcia2013x}. We justify using \code{relxilllp} because, in addition to being physically similar to \code{relxill}, it is parameterized by the AGN geometry, and therefore more diagnostic of it.     
With this model, we can quantify characteristics such as black hole spin (a=$cJ/GM^2$), coronal height above the accretion disk ($h$), ionization fraction ($log(\xi)$, where $\xi= L_{\rm{ion}}/(nr^2)$), reflection fraction (the ratio of flux incident on the disk to photons escaping to infinity, $f_{\rm{Refl}}$), iron abundance ($A_{\text{Fe}}$), and the inclination of the disk with respect to the line of sight ($i$). See \citealp{dauser2016normalizing} for discussion on the definition of the reflection fraction. 
The power-law continuum model accounts for a high energy cutoff such that $ F(E) \propto E ^{-\Gamma}e^{-E/E_{\rm{cut}}}$.  
Values for the best-fit \code{relxilllp} model parameters are presented in Table~\ref{tab_best2-10}.

\begin{table}[h!] 
\centering
\begin{tabular}{l l l}
\toprule
\multicolumn{3}{c}{Best-fit Model Excluding Soft Excess} \\ 

\hline
\hline
Model  & Component [units] & Value \\ 
\hline 
relxilllp & $h$ ($r_G$) &  <10 \\
 & $a$ & < 0.3  \\
  & $i$ ($^{\circ}$)& 37$^{+2}_{-2}$  \\ 
  & $\Gamma_{XMM}$ & 2.18$^{+0.02}_{-0.02}$ \\
    & $\Gamma_{\nstr}$ & 2.20$^{+0.04}_{-0.05}$  \\
 & $\text{log}~\xi_{ion}$ & 3.0$^{+0.1}_{-0.1}$ \\ 
   & $A_{\text{Fe}}$ ($A_{\text{Fe},\odot}$) & 3$^{+1}_{-1}$  \\
   & $E_{\rm{cut}}$ (keV) & > 200  \\
 & $f_{\rm{Refl}}$ & 1.8$^{+0.4}_{-0.3}$ \\
 \hline
tbnew & $N_{\text{H}}$ ($10^{22}$ cm$^{-2}$) & 0.126  \\
 \hline
 const & $C_{\rm{FPMA}}$ & 1.16$^{+0.07}_{-0.06}$  \\
  const & $C_{\rm{FPMB}}$ & 1.19$^{+0.07}_{-0.06}$  \\
 \hline
 \hline
 Fit & $\chi^2/\nu$ & $1487.72/1438 = 1.035$  \\
\bottomrule
\end{tabular}
\caption{Parameters obtained from spectral modeling of the time-integrated \xmm and \nstr spectra above 2 keV. All errors are at the 90\% confidence interval.} 
\label{tab_best2-10}
\end{table}

We measure a moderate coronal source height, a super-Solar iron abundance, and an ionization parameter and cutoff energy typical for AGN. This iron overabundance may be due to an intrinsically high iron abundance (the iron abundance increases towards the center of many galaxies, reaching values between 2 and 3 times Solar; e.g. \citealt{xu2018evolution}).
Interestingly, the measured spin is low compared to that of other AGN. 
\paragraph*{\textbf{Time-resolved Spectroscopy\\}}

\label{spc_bme}

We investigate differences between the high- and low-flux states during this observation, first quantifying changes in iron line width, luminosity, and flux, presented in Table~\ref{bmepars}. We begin by modeling the iron line with a redshifted Gaussian. The narrowing of the line during the low-flux state is apparent in the line widths reported in Table~\ref{bmepars}.

\begin{table*}[!ht] 
\centering
\begin{tabular}{l l l l l}
\toprule
\multicolumn{5}{c}{Model Parameters} \\
\midrule
\hline
Model  & Component & Pre-Dip & Low-Flux State & Post-Dip \\
\hline 
power law & $\Gamma_{XMM}$ & $2.04\pm0.03$ & $1.87\pm0.04$ & $2.12\pm0.03$ \\
& $\Gamma_{\nstr}$ & $2.18\pm0.04$ & $1.94\pm0.06$ & $2.04\pm0.04$ \\
zgaussian & $\sigma$ (keV) & $0.45\pm0.06$ & $0.23\pm0.06$ & $0.38\pm0.06$ \\ 
& $E_{\text{line}}$ (keV) & $6.54\pm0.05$ & $6.72\pm0.05$ & $6.84\pm0.05$ \\ 
constant & $C_{\rm{FPMA}}$ & $1.45\pm0.09$ & $0.68\pm0.07$ & $1.19\pm0.08$ \\
 & $C_{\rm{FPMB}}$ & $1.46\pm0.09$ & $0.67\pm0.07$ & $1.21\pm0.08$ \\
\hline
\hline
Luminosity & $L_{0.3-10~\text{keV}}$ (10$^{43}$ ergs s$^{-1}$) & 1.15  & 0.67 & 1.3 \\ 
Flux & $F_{2-10~\text{keV}}$ (10$^{-11}$ ergs cm$^{-2}$ s$^{-1}$) & 2.23 & 1.51 & 2.3 \\ 
 \hline
 \hline
 Fit & $\chi^2/\nu$ & $3303/3220 =1.02$ &   & \\ 
\bottomrule
\end{tabular}
\caption{Parameters obtained from spectral modeling of the time-resolved \xmm and \nstr spectra with \code{tbnew}*(\code{power law}+\code{zgaussian}) between 2$-$7 and 3$-$10 keV, respectively. 
The soft excess and Compton hump energies are omitted in order to measure the width of the iron line in each time bin during the observation. This allows for a simplistic characterization of the time-dependence of the prominent iron line profile between 6$-$7 keV as well as the power law continuum shape.}
\label{bmepars}
\end{table*}

During the low flux portion of the observation, we see an increase (hardening) in the continuum emission compared to the rest of the observation, accompanied by a decrease in the width of the iron line.

\begin{figure}[h!] 
\centering
\includegraphics[scale=0.47]{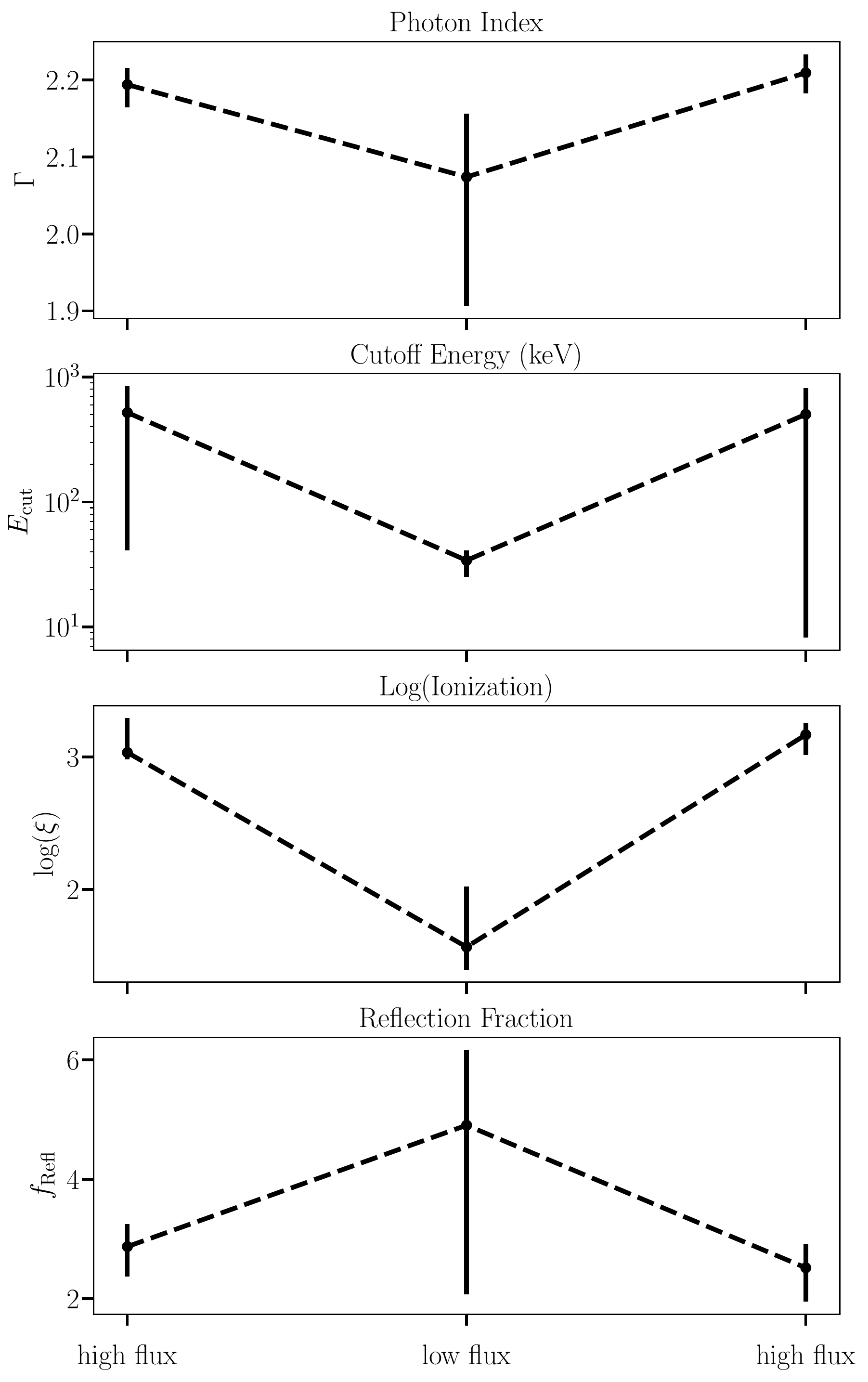}
\caption{ Parameters obtained from spectral modeling of the time-resolved \xmm and \nstr spectra in the energy range above 2 and 3 keV, respectively. Stated errors are at the 90\% level. Here we show all the parameters in the \code{relxilllp} model that are changing throughout the observation, all other parameters not shown are consistent with the time-integrated spectral modeling. }
\label{fig_bme2-7}
\end{figure}

We perform detailed modeling to understand the nature of these spectral changes, using the same best-fit model from the time-integrated analysis in Section~\ref{spc_int}, but now allowing the coronal height, ionization, reflection fraction, and continuum parameters to change throughout the three distinct flux states of the observation. Key parameters intrinsic to the black hole or disk geometry (and therefore not expected to change on 10-ks timescales), such as spin, inclination and iron abundance, are tied throughout the fitting. 
Values for variable model parameters are presented graphically in Figure~\ref{fig_bme2-7}. 
If the height of the corona is decreased during the low-flux state, this picture is consistent with the increase in reflection fraction we measure \citep{miniutti2004light}), as well as with the decrease in ionization (due to continuum attenuation by strong light bending, causing ionizing photons to fall beyond the horizon) and the cutoff energy (which is a proxy for coronal temperature). 
However, we cannot constrain a significant decrease in the coronal height throughout the observation; similarly, we measure no more than an indication of an increase in the reflection fraction.

\subsubsection{Modeling the Soft Excess}
\label{sec_softspc}
We observe a prominent and complex soft excess below 2 keV in this source. 
Starting with our initial model from Section~\ref{sec_fe}, an additional blackbody component is required to account for residuals at these soft energies. For this we utilize the disk multi-blackbody model \code{diskbb}, the spectrum of an accretion disk parameterized by the temperature at the inner-disk radius ($T_{\rm{in}}$, \citealp{mitsuda1984energy}). Local neutral absorption is also required in the form of a second \code{tbnew} multiplicative model component. 

\begin{figure}[h!] 
\centering
\includegraphics[scale=0.33]{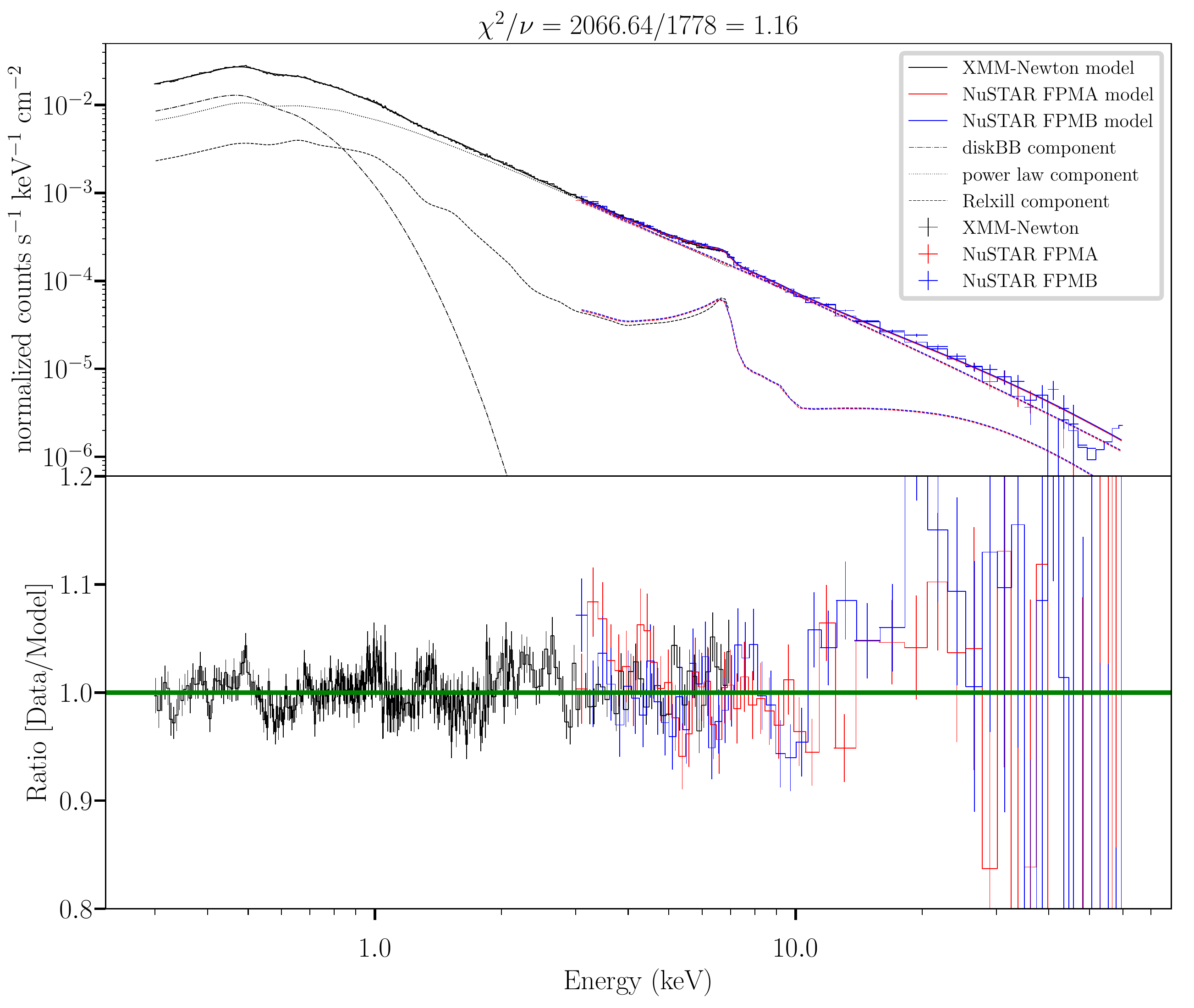}
\caption{Upper panel: \xmm (black) and \nstr (blue) spectra, this time including the energies of the soft excess (0.3$-$2 keV), fit to the model described in Table~\ref{tab_softfit} (power law + relativistically blurred reflection +  soft excess blackbody) convolved with local and Galactic neutral absorption. Data from both FPMs have been grouped for plotting purposes, and the data have been divided by the response effective area. Lower panel: Ratio residuals of \xmm and \nstr spectra compared to the best-fit model. }
\label{best_03-10}
\end{figure}

This best-fit model is represented in Figure~\ref{best_03-10} and associated parameter values are presented in Table~\ref{tab_softfit}. 
Residuals between 1.6 and 2.5 keV may be explained by miscalibration of the Au edge from the mirror and Si edges from the detector onboard \xmmns. 
Constraints on the coronal height and cutoff energy benefit from the additional data, and the spin remains strongly constrained to low or retrograde values. In the \code{relxilllp} model, the flag \code{fixReflFrac} = 2 will display the reflection fraction predicted from best-fit parameters in the lamppost geometry. Interpreting this as an upper limit accounts for different coronal geometries, e.g. an extended corona, for which $f_{\rm{Refl}}$ could be less. It can also exclude unphysical solutions of low spin and large reflection fractions. This maximum self-consistent reflection fraction is 1.11 (at $a=0.1$ and $h=5~r_{\rm{G}}$), consistent within 3 sigma of reflection fractions we report in Table~\ref{tab_softfit}, 1.16$-$1.32. The higher measured reflection fractions in Figure~\ref{fig_bme2-7} are in tension with theoretical predictions, though more robust fitting including the entire 0.3$-$60 keV energy spectra give a consistent picture. We measure an iron overabundance greater by a factor of 3 compared to the measurement of $A_{\rm{Fe}}$ from the spectrum above 2 keV, possibly due to a sensitivity of this parameter to the Fe L complex of lines near 1 keV \citep{turner1999arakelian}. 
The systematic residuals above 10 keV which were not seen when fitting only the hard X-ray spectra may similarly result from an overestimate of the Fe abundance when extending the fits to the soft band, as a consequence of trying to account for the soft excess with a reflection component only; however, fixing $A_{\rm{Fe}}$ to Solar resulted in significantly worse fits without describing the Compton hump well.

\begin{table}[h!]
\centering
\begin{tabular}{l l l}
\toprule
\multicolumn{3}{c}{Best-fit Model Including Soft Excess} \\ 
\midrule
\hline
Model      & Component [units]  & Value \\
\hline 
relxilllp & h ($r_G$) & < 5   \\
 & a & < 0.1 \\
  & $i$ ($^{\circ}$) & 35$^{+2}_{-1}$  \\  
  & $\Gamma_{XMM}$ & 2.16$^{+0.01}_{-0.01}$ \\
    & $\Gamma_{\nstr}$ & 2.14$^{+0.03}_{-0.02}$  \\
 & $\text{log}~\xi_{ion}$ & 3.01$^{+0.02}_{-0.02}$ \\ 
   & $A_{\text{Fe}}$ ($A_{\text{Fe},\odot}$)  & > 9.8 \\
   & $E_{cut}$ (keV) & 160$^{+80}_{-40}$ \\
 & $f_{\rm{Refl}}$ & 1.22$^{+0.10}_{-0.06}$ \\ 
 \hline
tbnew & $N_{\text{H}}$ ($10^{22}$ cm$^{-2}$) & 0.071$^{+0.005}_{-0.004}$  \\ 
 \hline 
diskbb & $T_{\rm{in}}$ (keV) & 0.125$^{+0.002}_{-0.003}$ \\ 
 \hline
 const & $C_{FPMA}$ & 1.07$^{+0.07}_{-0.01}$\\
  const & $C_{FPMB}$ & 1.09$^{+0.07}_{-0.01}$  \\
 \hline
 \hline
 Fit & $\chi^2/\nu$ & $2067/1779 = 1.16$ \\
\bottomrule
\end{tabular}
\caption{Parameters obtained from spectral modeling of the time-integrated \xmm and \nstr spectra above 0.3 keV.  }
\label{tab_softfit}
\end{table}

\subsection{Fourier-based Temporal Analysis} 
\label{tim}

\subsubsection{Time-integrated Lag Analysis}
Fourier-domain spectral-timing techniques to detect X-ray time lags (including X-ray reverberation mapping) consist of extracting the time delay from the phase given by the cross-spectrum of the hard (continuum) and soft band light curves. 
When combined with time-resolved spectroscopy, this provides a complementary picture of the source variability as well as the inner accretion-flow geometry. 

\begin{figure}[h!] 
\centering
\includegraphics[scale=0.3]{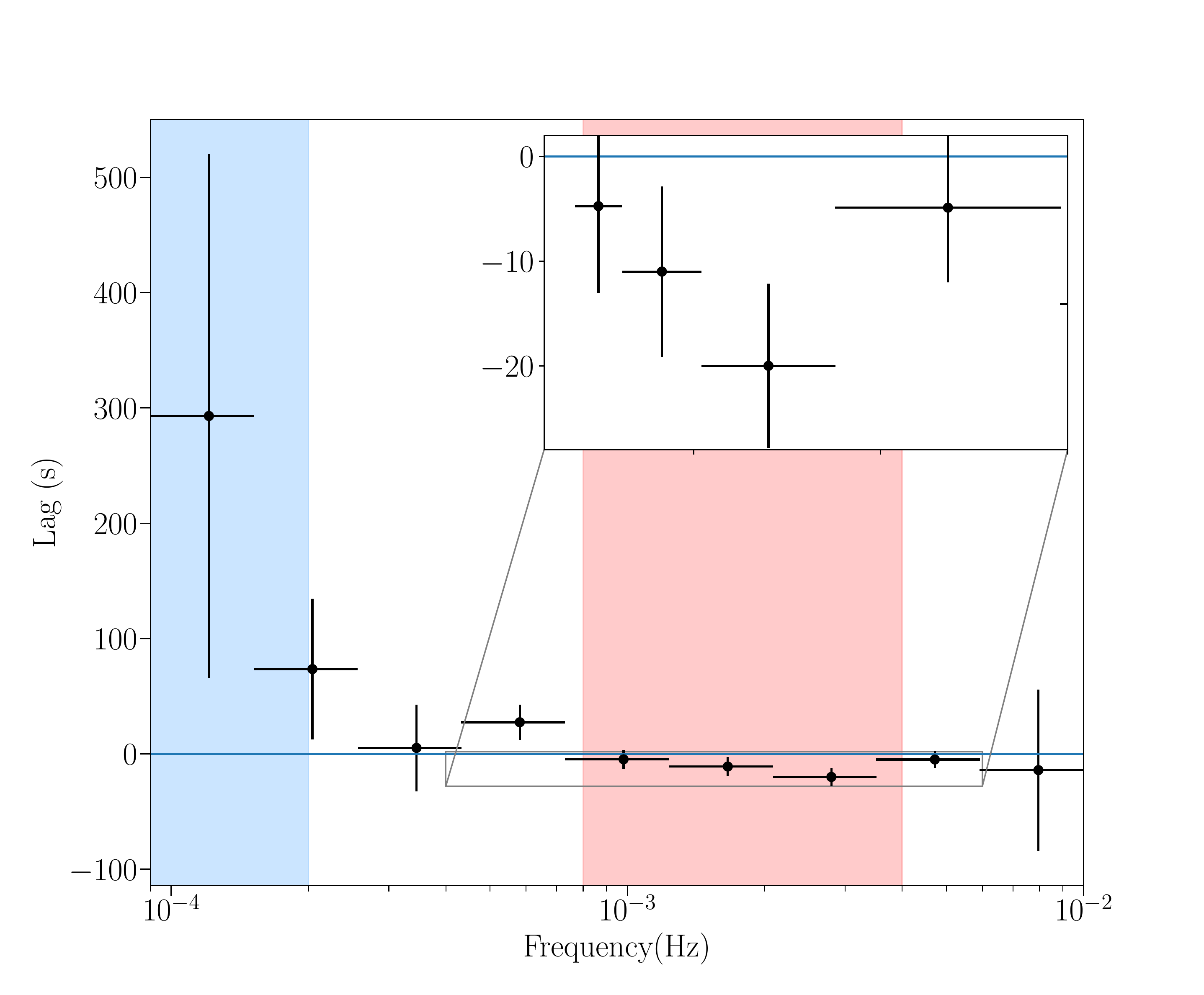}
\caption{ Lag frequency spectra of the \xmm \pn observation. The low-frequency variability in the 1$-$4 keV power-law-continuum-dominated band lags the reflection-dominated 0.3-1 keV energy band by hundreds of seconds (highlighted in blue), whereas it leads in the high-frequency regime (highlighted in red, zoomed in inset plot) by tens of seconds.} 
\label{lagfreq} 
\end{figure}

\begin{figure}[h!] 
\centering
\subfigure[]{
\includegraphics[scale=0.3]{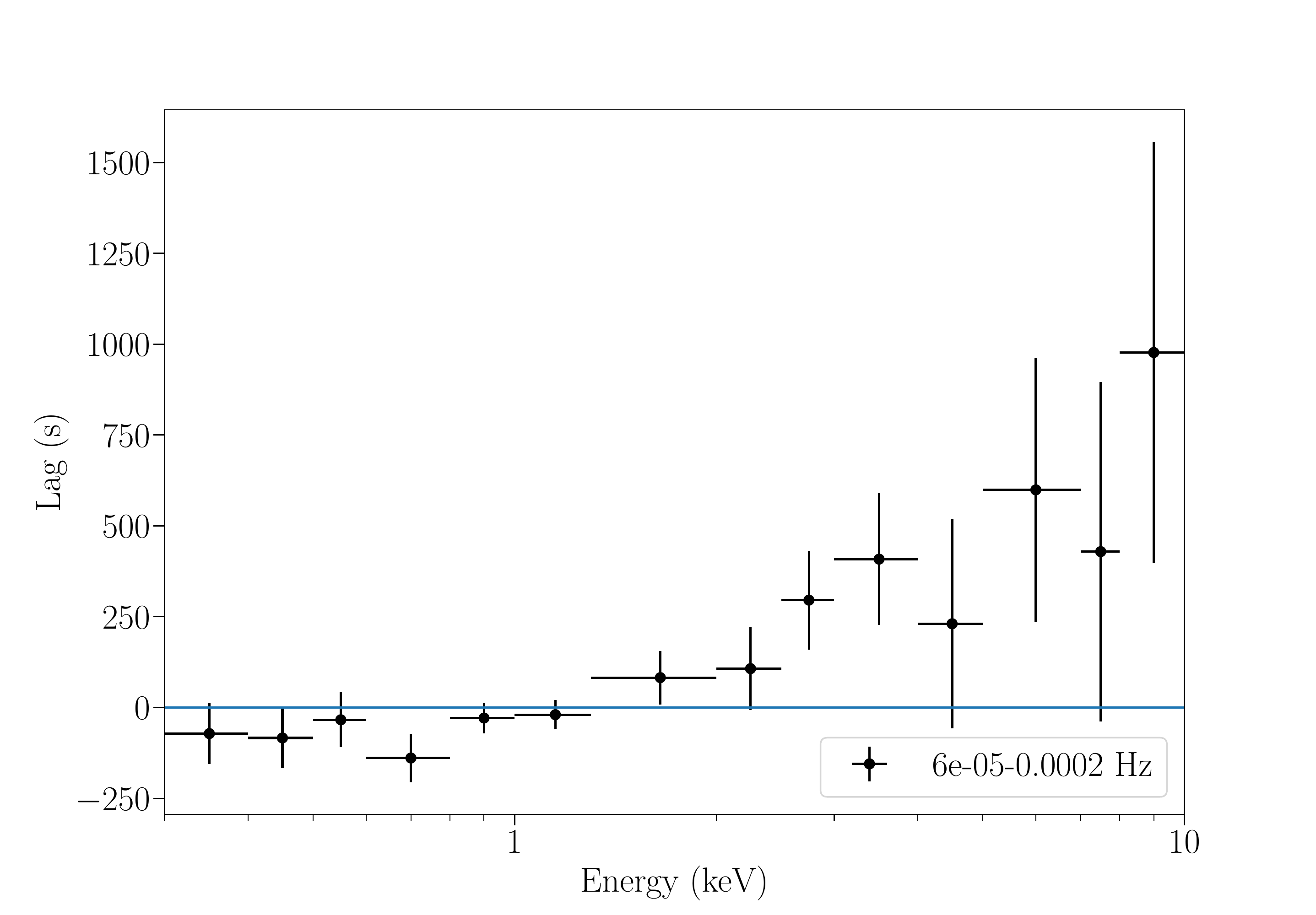}} 
\subfigure[]{
\includegraphics[scale=0.3]{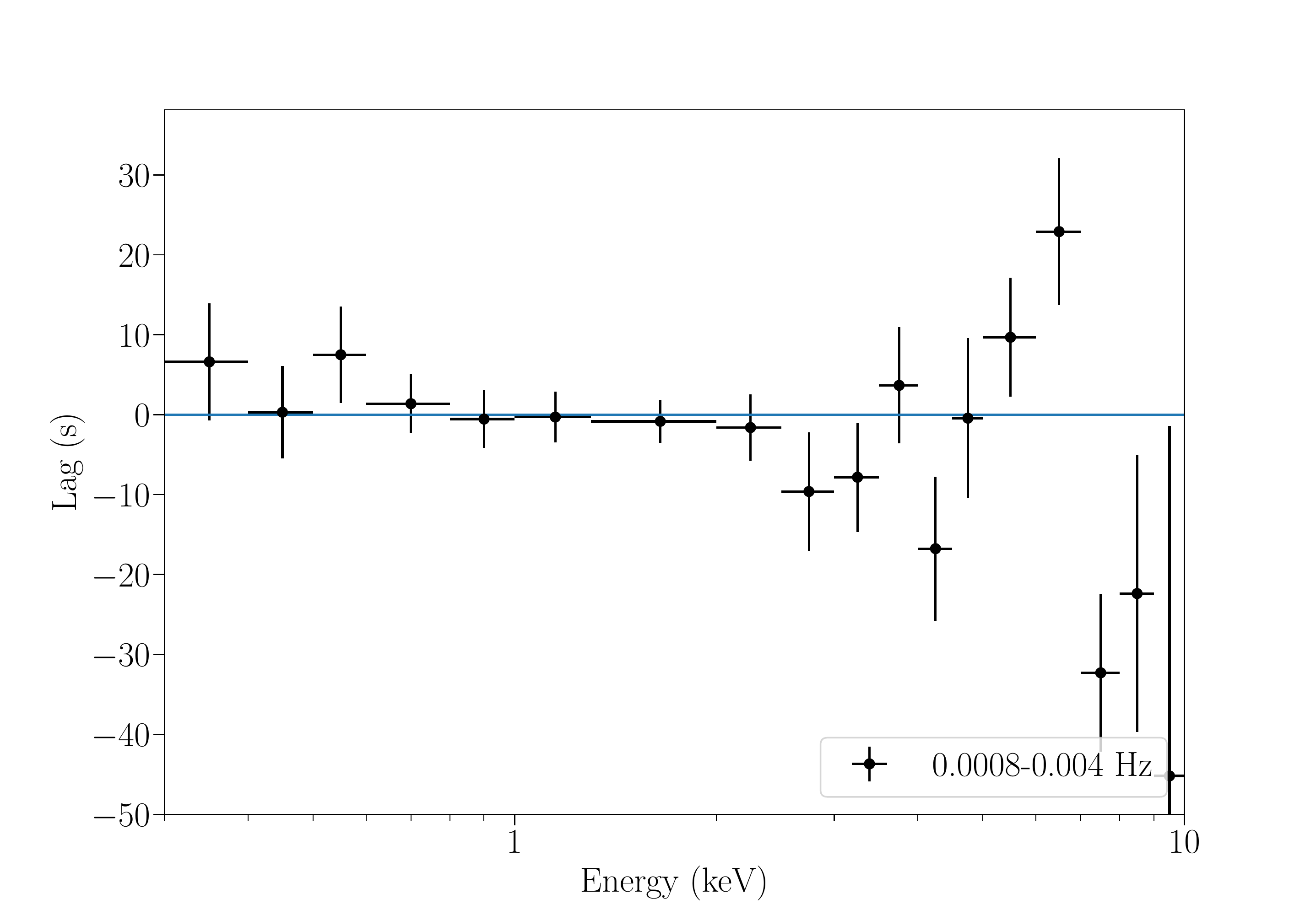}\label{lagen_hf}}
\caption{ Lag energy spectra over entire observation at the frequencies of the hard lag (a), and the soft lag (b). These frequency ranges are chosen based on the lag-frequency spectrum in Figure~\ref{lagfreq}, (highlighted in blue and red, respectively). In contrast to Figure~\ref{lagfreq}, the reference band of the lag is now taken to be all other energies between 0.3 and 10 keV (excluding the channel of interest), in order to maximize signal-to-noise. Note that errorbars extending beyond the plot frame are symmetric. An increase in the time delay can clearly be seen at the location of the Fe-K line. Lags with lower values precede higher-valued lags.} 
\label{lagen}
\end{figure}

For example, we can confirm the distance $h$ of the corona from the black hole obtained with spectral modeling with an independent estimate of $h$ using X-ray reverberation mapping. Following the methodology outlined in \citealp{uttley2014x}, we first plot the lag between emission components (distinguished based on their energies, 1$-$4 keV and 0.3$-$1 keV) decomposed into temporal frequencies, the lag frequency spectrum shown in Figure~\ref{lagfreq}. 
Both a low-frequency hard lag (where the continuum-dominated emission lags the soft reflection-dominated emission), and a high-frequency soft lag are seen in this observation. The hard lag is measured to be 300 $\pm$ 230 seconds at its maximum, and the soft lag (negative by convention) is measured at 20 $\pm$ 8 seconds. 
Given the frequency range where these time lags are taking place, we can compare an energy-resolved light curve to that of every other energy bin and thus build up a lag energy spectrum, shown in Figure~\ref{lagen} for both the hard and soft lags. 
There is a clear distinction between the shape of the lags at different temporal frequencies: the low-frequency lag between $\sim 10^{-5} - 10^{-4}$ Hz increases approximately log-linearly with energy, while at high frequencies ($\sim 10^{-3}$ Hz), there is structure in the lag tracing the Fe-K$\alpha$ emission line and agreement with a zero lag at lower energies despite a substantial soft excess. 
Given the statistically significant (compared to a simple power law) Gaussian-like feature at the energy of the broad iron line, we conclude Fe-K$\alpha$ reverberation is taking place during the observation. With this measurement, we can place \oneh in context with a number of other AGN for which the amplitude of the Fe-K$\alpha$ lag has been measured. A scaling between the amplitude of the lag and the black hole mass in a previous sample of Seyfert galaxies was reported by \citealp{de2013discovery}. \oneh fits nicely along the positive correlation, toward the low-mass, low-time-delay end of the relation. 

Before physical interpretation of this time-delay measurement, we must account for dilution by continuum photons in the soft band, using the fraction of continuum and reflection fluxes to estimate the dilution factor ($R=~F_{\rm{refl}}/F_{\rm{power~law}} \approx 0.3$) and reducing the derived continuum-to-reprocessor distance by the factor 
\begin{equation}
R_{0.3-1~{\rm keV}}/(1+R_{0.3-1~{\rm keV}}) - R_{1-4~{\rm keV}}/(1+R_{1-4~{\rm keV}})
\end{equation}
\citep{uttley2014x}. 
Relative fluxes were calculated using the \code{CFLUX} model in \code{XSPEC}.
Converting the magnitude of the high-frequency soft lag (20 seconds) into a light travel distance, (factoring in the large range of possible time delays as well as continuum dilution of the reflection spectrum) gives $ 9 \pm 4~r_{\rm G}$ for a black hole of mass $M = 3\times10^6 ~M_{\odot}$. 
This location from the accretion disk of the corona is consistent with our spectroscopic estimate of the coronal height $h \sim 2$-$5~r_G$, within a 95\% confidence limit. We emphasize that this is a zeroth order treatment of dilution, as these frequency-dependent time lags are in fact diluted only by the reflection fraction of the emission that is varying in the given frequency range, and that is correlated with the continuum emission. The blackbody component is not likely to vary on the same timescale as the power law and reflection components; therefore, we do not expect the blackbody component to contribute to the lags (or to the dilution).

\subsubsection{Time-resolved Lag Analysis}
\label{tim_bme}
\begin{figure}[h!]
\centering
\includegraphics[scale=0.3]{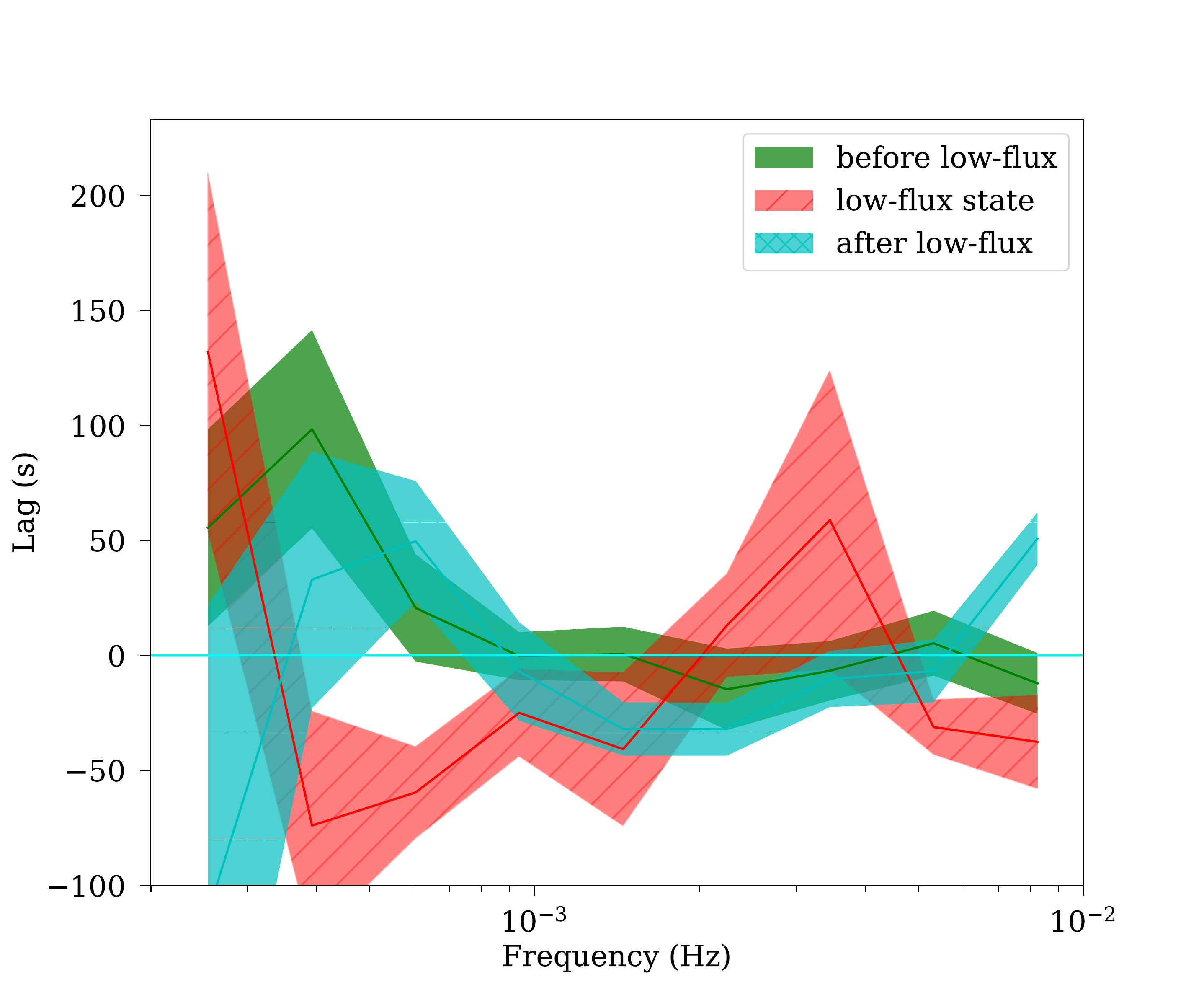}
\caption{Time-resolved lag-frequency spectrum showing the difference in the soft lag during the 20-ks low-flux state compared to the earlier and later stages of the observation.}
\label{overlags}
\end{figure}

In Section~\ref{lcs}, we have shown that the light curve of \oneh during this observation exhibits three different activity states: high-flux active variation, symmetric low emission and recovery, and a return to active variation. 
We explore the flux-dependence of these lags for each period of the observation, shown in Figure~\ref{overlags}. The low-flux state exhibits a very different lag-frequency spectrum than the earlier and later stages of the observation, with the soft lag shifting to lower frequencies.  The decreased reflection fraction during the high-flux states causes the delays to similarly be reduced during the early and late portions of the observation, because they are diluted by the continuum to the point that there is little-to-no significant lag.

\section{Discussion} 
\label{discuss}

\subsection{Is this an Eddington-limited AGN?}

This source has an Eddington ratio between 0.3 and 0.7 assuming $L_{\rm{bol}}/L_{2-10 \text{keV}} \sim 20-70 $, based on a simple power law relationship between the bolometric correction and Eddington ratio in AGN \citep{vasudevan2007piecing,kaspi2000reverberation}, where $L_{2-10 \text{keV}}=4.09\times10^{42}~\text{erg s}^{-1}$ for the redshift of this source, z = 0.0102. 
By this rough approximation, it is therefore toward the lower end of the distribution of accretion rates of NLS1s \citep{xu2012correlation}, although this estimate is lower than that measured by \citealp{malizia2008first} ($L_{\rm bol}/L_{Edd}\approx0.9$), who used the 1-100 keV luminosity as a proxy for bolometric luminosity, and the bolometric correction given by \citealp{risaliti2004panchromatic}, 4.2\%. 

Computing the Eddington ratio based on the known $\alpha_\text{OX}$-Eddington-ratio relation, where $\alpha_\text{OX}$ is the ratio of optical (2500 \AA) and X-ray (2 keV) spectral slopes, gives $L_{\rm bol}/L_{Edd} \approx$ 1.1.  
The mass estimate for \oneh inferred from line widths in \citealp{rodriguez2000visible} by \citealp{malizia2008first} has no error quoted, but we note that systematic uncertainties in mass estimates are typically $\sim$0.5 dex. 
The mean of these two measures gives a conservative, close-to-Eddington estimate of $L_{\rm bol}/L_{Edd} \approx$ 0.7, which places \oneh slightly below the average Eddington ratio for NLS1s given by \citealp{xu2012correlation}.

\subsection{Interpretation of Low-flux Event}

This observation reveals that \oneh shares many qualities with oft-studied NLS1s, such as strong reflection features, extreme variability, and a prominent soft excess. 
Its light curve reveals an event resulting in a depression of intensity, along with deviation from a single log-normal flux distribution, reminiscent of the flux dips of Fairall 9 \citep{lohfink2012black,lohfink2016rhythm}. 
Because the transit-like dip appears in the \nstr band, Compton-thin absorption cannot fully explain the decrease in flux. 
We show that any absorber would have to have a column density of $N_{\text{H}}\geq  10^{24}~\text{cm}^{-2}$ to explain dips in the 20$-$40 keV light curve, requiring a Compton-thick cloud. Since the flux does not fall to zero, the obscuring cloud would need to be of comparable size to the X-ray emitting region of the AGN ($R \sim \rm{few}\times GM/c^2\sim 10^{12}$ cm $\approx R_{\bigstar}$). Eclipsing the X-ray emitting region on a timescale of 20 ks implies an orbital velocity of 1000 km s$^{-1}$ and a radial distance of 10$^{16}$ cm, placing a cloud at the approximate location of the broad line region (BLR). 

A 1\% chance of a stellar occultation of the corona requires a star cluster with a total mass of 10$^8~M_\odot \approx 100 \times M_{\rm{BH}}$, a scenario which we conclude to be highly improbable, whereas a Compton-thick cloud requires merely $10^{-9}~M_{\rm{BH}}$. 

Interestingly, there is an additional (potentially log-normal) X-ray flux distribution associated with the dip, deviating from the bulk log-normal flux distribution in count rate only. This is consistent with a light-bending scenario, wherein the nature of the X-ray continuum (and therefore the shape of the distribution) would remain consistent, but the change in the average count rate of the distribution might correspond to a lower source height.

We determine that the nature of the symmetric dip seen in the light curve is not due to changes in accretion rate (as evidenced by lack of a similar flux decrease in the \textit{OM} \textit{UVW2} light curve). 

We combine detailed time-resolved spectral modeling with X-ray reverberation mapping to constrain source geometry, black hole properties, and emission/absorption processes dictating the nature of the variability of \onehns. The corona may have dropped down closer to the accretion disk causing the dip during this observation, a scenario which is consistent with the time-resolved X-ray spectroscopic measurements of increasing reflection fraction and decreasing coronal temperature and ionization during the low-flux state. 
We are limited largely by our understanding of the origin of the soft excess, an uncertainty which remains a source of systematic error for the spectral measurements. 
Studies of other NLS1s with large soft excesses result in conflicting best-fit models, a consequence of the absence of an agreed-upon universal shape for this spectral feature. 
If the soft excess is a result of blurred reflection off the inner accretion disk, then it should be able to be modeled fully by a relativistically convolved fluorescence model, (e.g. \citealp{gierlinski2004soft,crummy2006explanation}). 
The data, however, require an additional disk blackbody component. 
We estimate the Eddington ratio to be between 0.3 and 1.1, suggesting that \oneh has an accretion rate similar to measurements of other NLS1s. 

We report the detection of a relativistically-broadened Fe-K$\alpha$ emission line. The values of the spin derived from the best fit do not allow for differentiation between a Kerr and a Schwarzschild solution. The spin is low and iron abundance is consistently high (super-Solar) in both spectral models fits presented here (first excluding, then including the soft excess energies). The soft excess will be analyzed in more detail in a future paper describing analysis of the \textit{RGS} data. 

We measure a broad Fe-K soft lag at high temporal frequencies, confirming that this soft lag is due to reprocessing close to the central supermassive black hole. 
Note that although the high-frequency lag at the energy of the soft excess (below 2 keV as measured with respect to the 0.3$-$10 keV reference band\textemdash see Figure~\ref{lagen_hf}) is very small (< 10 seconds), the soft excess spectral feature is very prominent. This may suggest that the soft excess in this source contains another variable component besides reflection that dilutes the lag.

We also show that during the low-flux period of the observation, the low frequency hard lag appears to shift in frequency compared to the rest of the observation, which exhibits consistency before and after the dip. From the timing alone, we might deduce that this shift in frequency could be due to the corona becoming slightly more extended during the low-flux state. If the hard lag is diminished during the low-flux state (as in the anti-correlated flux dependence of the hard lag of NGC 4051 reported by \citealp{alston2013flux}), it is also possible that this apparent shift in frequency is the result of the persistent soft lag becoming visible at lower frequencies during that time. 

This scarcely-studied AGN \oneh may soon be considered one of the exemplary NLS1 prototypes, a new ideal laboratory for time-domain accretion physics.

\vspace{1em}
\footnotesize
\subsubsection*{Acknowledgements} 
\acknowledgments
We thank the anonymous referee for their helpful report. Based on observations obtained with XMM-Newton, an ESA science mission with instruments and contributions directly funded by ESA Member States and NASA. This work made use of data from the \nstr mission and is funded by \nstr Grant NNX15AV26G (PI: E. Kara). ACF acknowledges ERC Advanced Grant 340442. Thanks to M. Cole Miller, Chris Done, and J. Drew Hogg for useful discussions.

\bibliographystyle{aasjournal}
\bibliography{ref_2yp}

\begin{thebibliography}{}
\expandafter\ifx\csname natexlab\endcsname\relax\def\natexlab#1{#1}\fi
\providecommand{\url}[1]{\href{#1}{#1}}
\providecommand{\dodoi}[1]{doi:~\href{http://doi.org/#1}{\nolinkurl{#1}}}
\providecommand{\doeprint}[1]{\href{http://ascl.net/#1}{\nolinkurl{http://ascl.net/#1}}}
\providecommand{\doarXiv}[1]{\href{https://arxiv.org/abs/#1}{\nolinkurl{https://arxiv.org/abs/#1}}}

\bibitem[{Alston {et~al.}(2013)Alston, Vaughan, \& Uttley}]{alston2013flux}
Alston, W., Vaughan, S., \& Uttley, P. 2013, Monthly Notices of the Royal
  Astronomical Society, 435, 1511

\bibitem[{Arnaud {et~al.}(1985)Arnaud, Branduardi-Raymont, Culhane, Fabian,
  Hazard, McGlynn, Shafer, Tennant, \& Ward}]{arnaud1985exosat}
Arnaud, K., Branduardi-Raymont, G., Culhane, J., {et~al.} 1985, Monthly Notices
  of the Royal Astronomical Society, 217, 105

\bibitem[{{Arnaud}(1996)}]{1996ASPC..101...17A}
{Arnaud}, K.~A. 1996, in Astronomical Society of the Pacific Conference Series,
  Vol. 101, Astronomical Data Analysis Software and Systems V, ed. G.~H.
  {Jacoby} \& J.~{Barnes}, 17

\bibitem[{Baumgartner {et~al.}(2013)Baumgartner, Tueller, Markwardt, Skinner,
  Barthelmy, Mushotzky, Evans, \& Gehrels}]{baumgartner201370}
Baumgartner, W., Tueller, J., Markwardt, C., {et~al.} 2013, The Astrophysical
  Journal Supplement Series, 207, 19

\bibitem[{Bird {et~al.}(2007)Bird, Malizia, Bazzano, Barlow, Bassani, Hill,
  B{\'e}langer, Capitanio, Clark, Dean, {et~al.}}]{bird2007third}
Bird, A., Malizia, A., Bazzano, A., {et~al.} 2007, The Astrophysical Journal
  Supplement Series, 170, 175

\bibitem[{Boroson \& Green(1992)}]{boroson1992emission}
Boroson, T.~A., \& Green, R.~F. 1992, The Astrophysical Journal Supplement
  Series, 80, 109

\bibitem[{Brandt {et~al.}(1997)Brandt, Mathur, \& Elvis}]{brandt1997comparison}
Brandt, W., Mathur, S., \& Elvis, M. 1997, Monthly Notices of the Royal
  Astronomical Society, 285, L25

\bibitem[{{Cackett} {et~al.}(2013){Cackett}, {Fabian}, {Zogbhi}, {Kara},
  {Reynolds}, \& {Uttley}}]{2013ApJ...764L...9C}
{Cackett}, E.~M., {Fabian}, A.~C., {Zogbhi}, A., {et~al.} 2013, \apjl, 764, L9,
  \dodoi{10.1088/2041-8205/764/1/L9}

\bibitem[{Chatterjee {et~al.}(2009)Chatterjee, Marscher, Jorstad, Olmstead,
  McHardy, Aller, Aller, L{\"a}hteenm{\"a}ki, Tornikoski, Hovatta,
  {et~al.}}]{chatterjee2009disk}
Chatterjee, R., Marscher, A.~P., Jorstad, S.~G., {et~al.} 2009, The
  Astrophysical Journal, 704, 1689

\bibitem[{Condon {et~al.}(1998)Condon, Cotton, Greisen, Yin, Perley, Taylor, \&
  Broderick}]{condon1998nrao}
Condon, J., Cotton, W., Greisen, E., {et~al.} 1998, The Astronomical Journal,
  115, 1693

\bibitem[{Crummy {et~al.}(2006)Crummy, Fabian, Gallo, \&
  Ross}]{crummy2006explanation}
Crummy, J., Fabian, A., Gallo, L., \& Ross, R. 2006, Monthly Notices of the
  Royal Astronomical Society, 365, 1067

\bibitem[{Cutri {et~al.}(2003)Cutri, Skrutskie, Van~Dyk, Beichman, Carpenter,
  Chester, Cambresy, Evans, Fowler, Gizis, {et~al.}}]{cutri2003irsa}
Cutri, R., Skrutskie, M., Van~Dyk, S., {et~al.} 2003, NASA/IPAC Infrared
  Science Archive

\bibitem[{Dauser {et~al.}(2014)Dauser, Garc{\'\i}a, Parker, Fabian, \&
  Wilms}]{dauser2014role}
Dauser, T., Garc{\'\i}a, J., Parker, M., Fabian, A., \& Wilms, J. 2014, Monthly
  Notices of the Royal Astronomical Society: Letters, 444, L100

\bibitem[{Dauser {et~al.}(2016)Dauser, Garc{\'\i}a, Walton, Eikmann, Kallman,
  McClintock, \& Wilms}]{dauser2016normalizing}
Dauser, T., Garc{\'\i}a, J., Walton, D., {et~al.} 2016, Astronomy \&
  Astrophysics, 590, A76

\bibitem[{Dauser {et~al.}(2010)Dauser, Wilms, Reynolds, \&
  Brenneman}]{dauser2010broad}
Dauser, T., Wilms, J., Reynolds, C., \& Brenneman, L. 2010, Monthly Notices of
  the Royal Astronomical Society, 409, 1534

\bibitem[{Davison {et~al.}(1975)Davison, Culhane, Mitchell, \&
  Fabian}]{davison1975increase}
Davison, P., Culhane, J., Mitchell, R., \& Fabian, A. 1975, The Astrophysical
  Journal, 196, L23

\bibitem[{De~Marco {et~al.}(2013)De~Marco, Ponti, Cappi, Dadina, Uttley,
  Cackett, Fabian, \& Miniutti}]{de2013discovery}
De~Marco, B., Ponti, G., Cappi, M., {et~al.} 2013, Monthly Notices of the Royal
  Astronomical Society, 431, 2441

\bibitem[{Dewangan {et~al.}(2007)Dewangan, Griffiths, Dasgupta, \&
  Rao}]{dewangan2007investigation}
Dewangan, G., Griffiths, R., Dasgupta, S., \& Rao, A. 2007, The Astrophysical
  Journal, 671, 1284

\bibitem[{Dickey \& Lockman(1990)}]{dickey1990hi}
Dickey, J.~M., \& Lockman, F.~J. 1990, Annual review of astronomy and
  astrophysics, 28, 215

\bibitem[{Done {et~al.}(2012)Done, Davis, Jin, Blaes, \&
  Ward}]{done2012intrinsic}
Done, C., Davis, S., Jin, C., Blaes, O., \& Ward, M. 2012, Monthly Notices of
  the Royal Astronomical Society, 420, 1848

\bibitem[{Fabian {et~al.}(2015)Fabian, Lohfink, Kara, Parker, Vasudevan, \&
  Reynolds}]{fabian2015properties}
Fabian, A., Lohfink, A., Kara, E., {et~al.} 2015, Monthly Notices of the Royal
  Astronomical Society, 451, 4375

\bibitem[{Fabian \& Ross(2010)}]{fabian2010x}
Fabian, A., \& Ross, R. 2010, Space Science Reviews, 157, 167

\bibitem[{{Fabian} {et~al.}(2009){Fabian}, {Zoghbi}, {Ross}, {Uttley}, {Gallo},
  {Brandt}, {Blustin}, {Boller}, {Caballero-Garcia}, {Larsson}, {Miller},
  {Miniutti}, {Ponti}, {Reis}, {Reynolds}, {Tanaka}, \&
  {Young}}]{2009Natur.459..540F}
{Fabian}, A.~C., {Zoghbi}, A., {Ross}, R.~R., {et~al.} 2009, \nat, 459, 540,
  \dodoi{10.1038/nature08007}

\bibitem[{Gabriel {et~al.}(2004)Gabriel, Denby, Fyfe, Hoar, Ibarra, Ojero,
  Osborne, Saxton, Lammers, \& Vacanti}]{gabriel2004xmm}
Gabriel, C., Denby, M., Fyfe, D., {et~al.} 2004, in Astronomical Data Analysis
  Software and Systems (ADASS) XIII, Vol. 314, 759

\bibitem[{Galeev {et~al.}(1979)Galeev, Rosner, \&
  Vaiana}]{galeev1979structured}
Galeev, A., Rosner, R., \& Vaiana, G. 1979, The Astrophysical Journal, 229, 318

\bibitem[{Garcia {et~al.}(2013)Garcia, Dauser, Reynolds, Kallman, McClintock,
  Wilms, \& Eikmann}]{garcia2013x}
Garcia, J., Dauser, T., Reynolds, C., {et~al.} 2013, The Astrophysical Journal,
  768, 146

\bibitem[{Garc{\'\i}a {et~al.}(2011)Garc{\'\i}a, Kallman, \&
  Mushotzky}]{garcia2011x}
Garc{\'\i}a, J., Kallman, T., \& Mushotzky, R. 2011, The Astrophysical Journal,
  731, 131

\bibitem[{Garc{\'\i}a {et~al.}(2014)Garc{\'\i}a, Dauser, Lohfink, Kallman,
  Steiner, McClintock, Brenneman, Wilms, Eikmann, Reynolds,
  {et~al.}}]{garcia2014improved}
Garc{\'\i}a, J., Dauser, T., Lohfink, A., {et~al.} 2014, The Astrophysical
  Journal, 782, 76

\bibitem[{Gaskell(2004)}]{gaskell2004lognormal}
Gaskell, C.~M. 2004, The Astrophysical Journal Letters, 612, L21

\bibitem[{George \& Fabian(1991)}]{george1991x}
George, I., \& Fabian, A. 1991, Monthly Notices of the Royal Astronomical
  Society, 249, 352

\bibitem[{Gierli{\'n}ski \& Done(2004)}]{gierlinski2004soft}
Gierli{\'n}ski, M., \& Done, C. 2004, Monthly Notices of the Royal Astronomical
  Society, 349, L7

\bibitem[{Gondek {et~al.}(1996)Gondek, Zdziarski, Johnson, George,
  McNaron-Brown, Magdziarz, Smith, \& Gruber}]{gondek1996average}
Gondek, D., Zdziarski, A.~A., Johnson, W.~N., {et~al.} 1996, Monthly Notices of
  the Royal Astronomical Society, 282, 646

\bibitem[{Goodrich(1989)}]{goodrich1989spectropolarimetry}
Goodrich, R.~W. 1989, The Astrophysical Journal, 342, 224

\bibitem[{Haardt \& Maraschi(1991)}]{haardt1991two}
Haardt, F., \& Maraschi, L. 1991, The Astrophysical Journal, 380, L51

\bibitem[{Haardt \& Maraschi(1993)}]{haardt1993x}
---. 1993, The Astrophysical Journal, 413, 507

\bibitem[{Harrison {et~al.}(2013)Harrison, Craig, Christensen, Hailey, Zhang,
  Boggs, Stern, Cook, Forster, Giommi, {et~al.}}]{harrison2013nuclear}
Harrison, F.~A., Craig, W.~W., Christensen, F.~E., {et~al.} 2013, The
  Astrophysical Journal, 770, 103

\bibitem[{Hiroi {et~al.}(2013)Hiroi, Ueda, Hayashida, Shidatsu, Sato, Kawamuro,
  Sugizaki, Nakahira, Serino, Kawai, {et~al.}}]{hiroi201337}
Hiroi, K., Ueda, Y., Hayashida, M., {et~al.} 2013, The Astrophysical Journal
  Supplement Series, 207, 36

\bibitem[{{Ives} {et~al.}(1976){Ives}, {Sanford}, \&
  {Penston}}]{1976ApJ...207L.159I}
{Ives}, J.~C., {Sanford}, P.~W., \& {Penston}, M.~V. 1976, \apjl, 207, L159,
  \dodoi{10.1086/182203}

\bibitem[{Jin {et~al.}(2015)Jin, Done, \& Ward}]{jin2015strong}
Jin, C., Done, C., \& Ward, M. 2015, Monthly Notices of the Royal Astronomical
  Society, 455, 691

\bibitem[{Jourdain {et~al.}(1992)Jourdain, Bassani, Bouchet, Mandrou, Ballet,
  Lebrun, Paul, Laurent, Churazov, Gilfanov, {et~al.}}]{jourdain1992sigma}
Jourdain, E., Bassani, L., Bouchet, L., {et~al.} 1992, Astronomy and
  Astrophysics, 256, L38

\bibitem[{Kalberla {et~al.}(2005)Kalberla, Burton, Hartmann, Arnal, Bajaja,
  Morras, \& P{\"o}ppel}]{kalberla2005leiden}
Kalberla, P.~M., Burton, W., Hartmann, D., {et~al.} 2005, Astronomy \&
  Astrophysics, 440, 775

\bibitem[{{Kara} {et~al.}(2016){Kara}, {Alston}, {Fabian}, {Cackett}, {Uttley},
  {Reynolds}, \& {Zoghbi}}]{kara2016global}
{Kara}, E., {Alston}, W.~N., {Fabian}, A.~C., {et~al.} 2016, Mon. Not. R.
  Astron. Soc., 462, 511, \dodoi{10.1093/mnras/stw1695}

\bibitem[{{Kara} {et~al.}(2014{\natexlab{a}}){Kara}, {Cackett}, {Fabian},
  {Reynolds}, \& {Uttley}}]{2014MNRAS.439L..26K}
{Kara}, E., {Cackett}, E.~M., {Fabian}, A.~C., {Reynolds}, C., \& {Uttley}, P.
  2014{\natexlab{a}}, \mnras, 439, L26, \dodoi{10.1093/mnrasl/slt173}

\bibitem[{{Kara} {et~al.}(2013){Kara}, {Fabian}, {Cackett}, {Uttley},
  {Wilkins}, \& {Zoghbi}}]{2013MNRAS.434.1129K}
{Kara}, E., {Fabian}, A.~C., {Cackett}, E.~M., {et~al.} 2013, \mnras, 434,
  1129, \dodoi{10.1093/mnras/stt1055}

\bibitem[{{Kara} {et~al.}(2014{\natexlab{b}}){Kara}, {Fabian}, {Marinucci},
  {Matt}, {Parker}, {Alston}, {Brenneman}, {Cackett}, \&
  {Miniutti}}]{2014MNRAS.445...56K}
{Kara}, E., {Fabian}, A.~C., {Marinucci}, A., {et~al.} 2014{\natexlab{b}},
  \mnras, 445, 56, \dodoi{10.1093/mnras/stu1750}

\bibitem[{{Kara} {et~al.}(2015){Kara}, {Zoghbi}, {Marinucci}, {Walton},
  {Fabian}, {Risaliti}, {Boggs}, {Christensen}, {Fuerst}, {Hailey}, {Harrison},
  {Matt}, {Parker}, {Reynolds}, {Stern}, \& {Zhang}}]{2015MNRAS.446..737K}
{Kara}, E., {Zoghbi}, A., {Marinucci}, A., {et~al.} 2015, \mnras, 446, 737,
  \dodoi{10.1093/mnras/stu2136}

\bibitem[{Kaspi {et~al.}(2000)Kaspi, Smith, Netzer, Maoz, Jannuzi, \&
  Giveon}]{kaspi2000reverberation}
Kaspi, S., Smith, P.~S., Netzer, H., {et~al.} 2000, The Astrophysical Journal,
  533, 631

\bibitem[{Kataoka {et~al.}(2007)Kataoka, Reeves, Iwasawa, Markowitz, Mushotzky,
  Arimoto, Takahashi, Tsubuku, Ushio, Watanabe, {et~al.}}]{kataoka2007probing}
Kataoka, J., Reeves, J.~N., Iwasawa, K., {et~al.} 2007, Publications of the
  Astronomical Society of Japan, 59, 279

\bibitem[{Komossa {et~al.}(2006)Komossa, Voges, Xu, Mathur, Adorf, Lemson,
  Duschl, \& Grupe}]{komossa2006radio}
Komossa, S., Voges, W., Xu, D., {et~al.} 2006, The Astronomical Journal, 132,
  531

\bibitem[{Kotov {et~al.}(2001)Kotov, Churazov, \& Gilfanov}]{kotov2001x}
Kotov, O., Churazov, E., \& Gilfanov, M. 2001, Monthly Notices of the Royal
  Astronomical Society, 327, 799

\bibitem[{{Leighly} {et~al.}(1996){Leighly}, {Mushotzky}, {Yaqoob}, {Kunieda},
  \& {Edelson}}]{leighly1996x}
{Leighly}, K.~M., {Mushotzky}, R.~F., {Yaqoob}, T., {Kunieda}, H., \&
  {Edelson}, R. 1996, \apj, 469, 147, \dodoi{10.1086/177767}

\bibitem[{Lohfink {et~al.}(2016)Lohfink, Reynolds, Pinto, Alston, Boggs,
  Christensen, Craig, Fabian, Hailey, Harrison, {et~al.}}]{lohfink2016rhythm}
Lohfink, A., Reynolds, C., Pinto, C., {et~al.} 2016, The Astrophysical Journal,
  821, 11

\bibitem[{Lohfink {et~al.}(2012)Lohfink, Reynolds, Miller, Brenneman,
  Mushotzky, Nowak, \& Fabian}]{lohfink2012black}
Lohfink, A.~M., Reynolds, C.~S., Miller, J.~M., {et~al.} 2012, The
  Astrophysical Journal, 758, 67

\bibitem[{Malizia {et~al.}(2008)Malizia, Bassani, Bird, Landi, Masetti,
  De~Rosa, Panessa, Molina, Dean, Perri, {et~al.}}]{malizia2008first}
Malizia, A., Bassani, L., Bird, A., {et~al.} 2008, Monthly Notices of the Royal
  Astronomical Society, 389, 1360

\bibitem[{Matsuoka {et~al.}(1990)Matsuoka, Piro, Yamauchi, \&
  Murakami}]{matsuoka1990x}
Matsuoka, M., Piro, L., Yamauchi, M., \& Murakami, T. 1990, The Astrophysical
  Journal, 361, 440

\bibitem[{Merloni \& Fabian(2001)}]{merloni2001accretion}
Merloni, A., \& Fabian, A. 2001, Monthly Notices of the Royal Astronomical
  Society, 321, 549

\bibitem[{Miniutti \& Fabian(2004)}]{miniutti2004light}
Miniutti, G., \& Fabian, A. 2004, Monthly Notices of the Royal Astronomical
  Society, 349, 1435

\bibitem[{Mitsuda {et~al.}(1984)Mitsuda, Inoue, Koyama, Makishima, Matsuoka,
  Ogawara, Suzuki, Tanaka, Shibazaki, \& Hirano}]{mitsuda1984energy}
Mitsuda, K., Inoue, H., Koyama, K., {et~al.} 1984, Publications of the
  Astronomical Society of Japan, 36, 741

\bibitem[{Molkov {et~al.}(2004)Molkov, Cherepashchuk, Lutovinov, Revnivtsev,
  Postnov, \& Sunyaev}]{molkov2004hard}
Molkov, S., Cherepashchuk, A., Lutovinov, A., {et~al.} 2004, Astronomy Letters,
  30, 534

\bibitem[{Nagao {et~al.}(2001)Nagao, Murayama, \& Taniguchi}]{nagao2001narrow}
Nagao, T., Murayama, T., \& Taniguchi, Y. 2001, The Astrophysical Journal, 546,
  744

\bibitem[{Nandra \& Pounds(1994)}]{nandra1994ginga}
Nandra, K., \& Pounds, K. 1994, Monthly Notices of the Royal Astronomical
  Society, 268, 405

\bibitem[{Nied{\'z}wiecki \& Zdziarski(2018)}]{niedzwiecki2018lamppost}
Nied{\'z}wiecki, A., \& Zdziarski, A.~A. 2018, Monthly Notices of the Royal
  Astronomical Society

\bibitem[{Osterbrock \& Pogge(1985)}]{osterbrock1985spectra}
Osterbrock, D.~E., \& Pogge, R.~W. 1985, The Astrophysical Journal, 297, 166

\bibitem[{Page {et~al.}(2004)Page, Schartel, Turner, \& O'Brien}]{page2004xmm}
Page, K.~L., Schartel, N., Turner, M.~J., \& O'Brien, P.~T. 2004, Monthly
  Notices of the Royal Astronomical Society, 352, 523

\bibitem[{Panessa {et~al.}(2011)Panessa, De~Rosa, Bassani, Bazzano, Bird,
  Landi, Malizia, Miniutti, Molina, \& Ubertini}]{panessa2011narrow}
Panessa, F., De~Rosa, A., Bassani, L., {et~al.} 2011, Monthly Notices of the
  Royal Astronomical Society, 417, 2426

\bibitem[{Piconcelli {et~al.}(2005)Piconcelli, Jimenez-Bail{\'o}n, Guainazzi,
  Schartel, Rodr{\'\i}guez-Pascual, \& Santos-Lle{\'o}}]{piconcelli2005xmm}
Piconcelli, E., Jimenez-Bail{\'o}n, E., Guainazzi, M., {et~al.} 2005, Astronomy
  \& Astrophysics, 432, 15

\bibitem[{Piro {et~al.}(1990)Piro, Yamauchi, \& Matsuoka}]{piro1990x}
Piro, L., Yamauchi, M., \& Matsuoka, M. 1990, The Astrophysical Journal, 360,
  L35

\bibitem[{Ponti {et~al.}(2012)Ponti, Papadakis, Bianchi, Guainazzi, Matt,
  Uttley, \& Bonilla}]{ponti2012caixa}
Ponti, G., Papadakis, I., Bianchi, S., {et~al.} 2012, Astronomy \&
  Astrophysics, 542, A83

\bibitem[{Porquet {et~al.}(2004)Porquet, Reeves, O'Brien, \&
  Brinkmann}]{porquet2004xmm}
Porquet, D., Reeves, J., O'Brien, P., \& Brinkmann, W. 2004, Astronomy \&
  Astrophysics, 422, 85

\bibitem[{Pounds {et~al.}(1995)Pounds, Done, \& Osborne}]{pounds1995re}
Pounds, K., Done, C., \& Osborne, J. 1995, Monthly Notices of the Royal
  Astronomical Society, 277, L5

\bibitem[{Pounds {et~al.}(1990)Pounds, Nandra, Stewart, George, \&
  Fabian}]{pounds1990x}
Pounds, K., Nandra, K., Stewart, G., George, I., \& Fabian, A. 1990, Nature,
  344, 132

\bibitem[{Reis \& Miller(2013)}]{reis2013size}
Reis, R., \& Miller, J. 2013, The Astrophysical Journal Letters, 769, L7

\bibitem[{Risaliti \& Elvis(2004)}]{risaliti2004panchromatic}
Risaliti, G., \& Elvis, M. 2004, in Supermassive Black Holes in the Distant
  Universe (Springer), 187--224

\bibitem[{Rodriguez {et~al.}(2008)Rodriguez, Tomsick, \&
  Chaty}]{rodriguez2008swift}
Rodriguez, J., Tomsick, J., \& Chaty, S. 2008, Astronomy \& Astrophysics, 482,
  731

\bibitem[{Rodr{\'\i}guez-Ardila {et~al.}(2000)Rodr{\'\i}guez-Ardila, Pastoriza,
  \& Donzelli}]{rodriguez2000visible}
Rodr{\'\i}guez-Ardila, A., Pastoriza, M.~G., \& Donzelli, C.~J. 2000, The
  Astrophysical Journal Supplement Series, 126, 63

\bibitem[{Sanfrutos {et~al.}(2013)Sanfrutos, Miniutti, Ag{\'\i}s-Gonz{\'a}lez,
  Fabian, Miller, Panessa, \& Zoghbi}]{sanfrutos2013size}
Sanfrutos, M., Miniutti, G., Ag{\'\i}s-Gonz{\'a}lez, B., {et~al.} 2013, Monthly
  Notices of the Royal Astronomical Society, 436, 1588

\bibitem[{Schurch \& Done(2006)}]{schurch2006failed}
Schurch, N.~J., \& Done, C. 2006, Monthly Notices of the Royal Astronomical
  Society, 371, 81

\bibitem[{Singh {et~al.}(1985)Singh, Garmire, \& Nousek}]{singh1985observation}
Singh, K., Garmire, G., \& Nousek, J. 1985, The Astrophysical Journal, 297, 633

\bibitem[{Str{\"u}der {et~al.}(2001)Str{\"u}der, Briel, Dennerl, Hartmann,
  Kendziorra, Meidinger, Pfeffermann, Reppin, Aschenbach, Bornemann,
  {et~al.}}]{struder2001european}
Str{\"u}der, L., Briel, U., Dennerl, K., {et~al.} 2001, Astronomy \&
  Astrophysics, 365, L18

\bibitem[{Sunyaev \& Titarchuk(1980)}]{sunyaev1980comptonization}
Sunyaev, R., \& Titarchuk, L. 1980, Astronomy and Astrophysics, 86, 121

\bibitem[{Taylor \& Reynolds(2018)}]{taylor2018exploring}
Taylor, C., \& Reynolds, C.~S. 2018, The Astrophysical Journal, 855, 120

\bibitem[{Treves {et~al.}(1990)Treves, Bonelli, Chiappetti, Falomo, Maraschi,
  Tagliaferri, \& Tanzi}]{treves1990x}
Treves, A., Bonelli, G., Chiappetti, L., {et~al.} 1990, The Astrophysical
  Journal, 359, 98

\bibitem[{Turner {et~al.}(1999)Turner, George, \& Netzer}]{turner1999arakelian}
Turner, T., George, I., \& Netzer, H. 1999, The Astrophysical Journal, 526, 52

\bibitem[{Turner \& Pounds(1989)}]{turner1989exosat}
Turner, T., \& Pounds, K. 1989, Monthly Notices of the Royal Astronomical
  Society, 240, 833

\bibitem[{Uttley {et~al.}(2014)Uttley, Cackett, Fabian, Kara, \&
  Wilkins}]{uttley2014x}
Uttley, P., Cackett, E., Fabian, A., Kara, E., \& Wilkins, D. 2014, The
  Astronomy and Astrophysics Review, 22, 72

\bibitem[{Vasudevan \& Fabian(2007)}]{vasudevan2007piecing}
Vasudevan, R.~V., \& Fabian, A. 2007, Monthly Notices of the Royal Astronomical
  Society, 381, 1235

\bibitem[{{Verner} {et~al.}(1996){Verner}, {Ferland}, {Korista}, \&
  {Yakovlev}}]{verner1996atomic}
{Verner}, D.~A., {Ferland}, G.~J., {Korista}, K.~T., \& {Yakovlev}, D.~G. 1996,
  \apj, 465, 487, \dodoi{10.1086/177435}

\bibitem[{Wilkins \& Gallo(2015)}]{wilkins2015driving}
Wilkins, D., \& Gallo, L.~C. 2015, Monthly Notices of the Royal Astronomical
  Society, 449, 129

\bibitem[{Wilms {et~al.}(2000)Wilms, Allen, \& McCray}]{wilms2000absorption}
Wilms, J., Allen, A., \& McCray, R. 2000, The Astrophysical Journal, 542, 914

\bibitem[{{Winkler} \& {White}(1975)}]{1975ApJ...199L.139W}
{Winkler}, Jr., P.~F., \& {White}, A.~E. 1975, \apjl, 199, L139,
  \dodoi{10.1086/181868}

\bibitem[{Xu {et~al.}(2012)Xu, Komossa, Zhou, Lu, Li, Grupe, Wang, \&
  Yuan}]{xu2012correlation}
Xu, D., Komossa, S., Zhou, H., {et~al.} 2012, The Astronomical Journal, 143, 83

\bibitem[{Xu {et~al.}(2018)Xu, Bian, Shen, Zuo, Fan, \& Zhu}]{xu2018evolution}
Xu, F., Bian, F., Shen, Y., {et~al.} 2018, Monthly Notices of the Royal
  Astronomical Society

\bibitem[{{Zoghbi} {et~al.}(2012){Zoghbi}, {Fabian}, {Reynolds}, \&
  {Cackett}}]{2012MNRAS.422..129Z}
{Zoghbi}, A., {Fabian}, A.~C., {Reynolds}, C.~S., \& {Cackett}, E.~M. 2012,
  \mnras, 422, 129, \dodoi{10.1111/j.1365-2966.2012.20587.x}

\end{thebibliography}

\end{document}